\documentclass[a4paper]{article}

\usepackage{amsmath,amstext,amsgen,amsbsy,amsopn,amsfonts,amssymb}
\usepackage{easybmat}
\usepackage{graphics}
\usepackage[pdftex]{graphicx}
\usepackage{epsfig}
\usepackage{amsthm}
\usepackage{bm}
\newtheorem{theorem}{Theorem}
\newtheorem{lemma}{Lemma}

\begin{document}
\title{Classification under Data Contamination with Application to Remote Sensing Image Mis-registration}

\author{
Donghui Yan$^{\dag\P}$, Peng Gong$^{\S\P}$, Aiyou Chen$^{\ddag}$,
Liheng Zhong$^{\S\P}$
\\\\
$^{\dag}$Department of Statistics\\
$^{\S}$Department of Environmental Science, Policy and Management\\
$^\P$University of California, Berkeley, CA 94720\\
$^{\ddag}$Google, Mountain View, CA 94043\footnote{The authors can
be contacted at: dhyan@berkeley.edu, penggong@berkeley.edu,
Aiyouchen@google.com, lihengzhong@berkeley.edu. This work was
partially done while AC was at Bell Labs, Murray Hill, NJ.} \\
}

\date{}
\maketitle

\begin{abstract}
This work is motivated by the problem of image mis-registration in
remote sensing and we are interested in determining the resulting
loss in the accuracy of pattern classification. A statistical
formulation is given where we propose to use data contamination to
model the phenomenon of image mis-registration. This model is widely
applicable to many other types of errors as well, for example,
measurement errors and gross errors etc. The impact of data
contamination on classification is studied under a statistical
learning theoretical framework. A closed-form asymptotic bound is
established for the resulting loss in classification accuracy, which
is less than $\epsilon/(1-\epsilon)$ for data contamination of an
amount of $\epsilon$. Our bound is sharper than similar bounds in
the domain adaptation literature and, unlike such bounds, it applies 
to classifiers with an infinite VC dimension. Extensive
simulations have been conducted on both synthetic and real datasets
under various types of data contaminations, including label
flipping, feature swapping and the replacement of feature values
with data generated from a random source such as a Gaussian or
Cauchy distribution. Our simulation results show that the bound we
derive is fairly tight.
\end{abstract}
\maketitle

\section{Introduction}
\label{section:Introduction} A motivating example of this work is
the problem of image mis-registration which occurs almost
ubiquitously in remote sensing. Image mis-registration refers to the
phenomenon where the image of interest is mapped or aligned to a
wrong position.  This is usually caused by errors in the image or
data acquisition device or the inaccuracy of the underlying mapping
algorithms which try to map data collected at different scales, at
different times, or taken from different angles.
Figure~\ref{figure:RSimage} below illustrates an instance of image
mis-registration where the image is tilted and then shifted by a
small amount. 
\begin{figure}[ht]
\centering
\begin{center}
\hspace{0cm}
\includegraphics*[scale=0.28,clip]{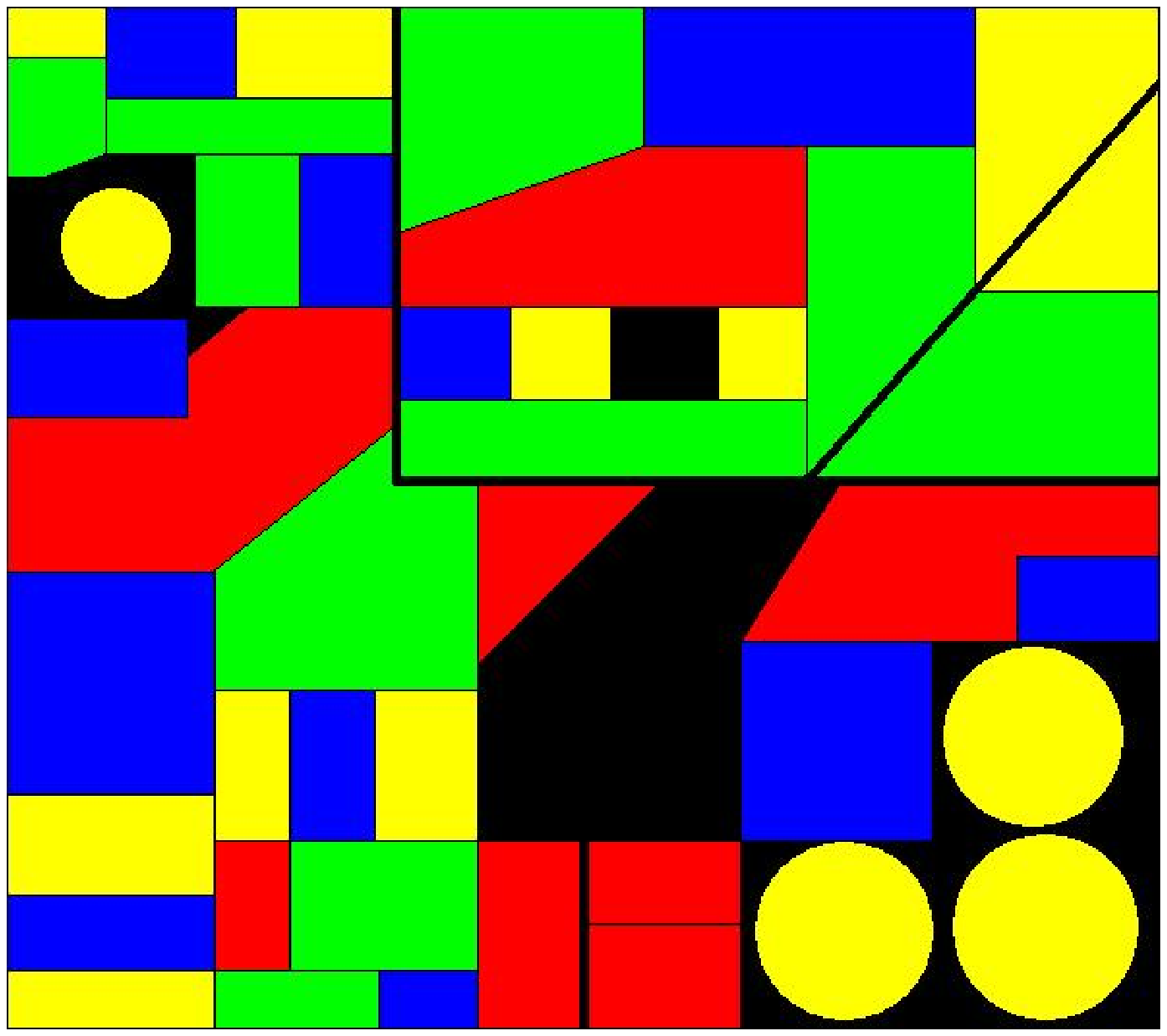}
\hspace{0.1in}
\includegraphics*[scale=0.28,clip]{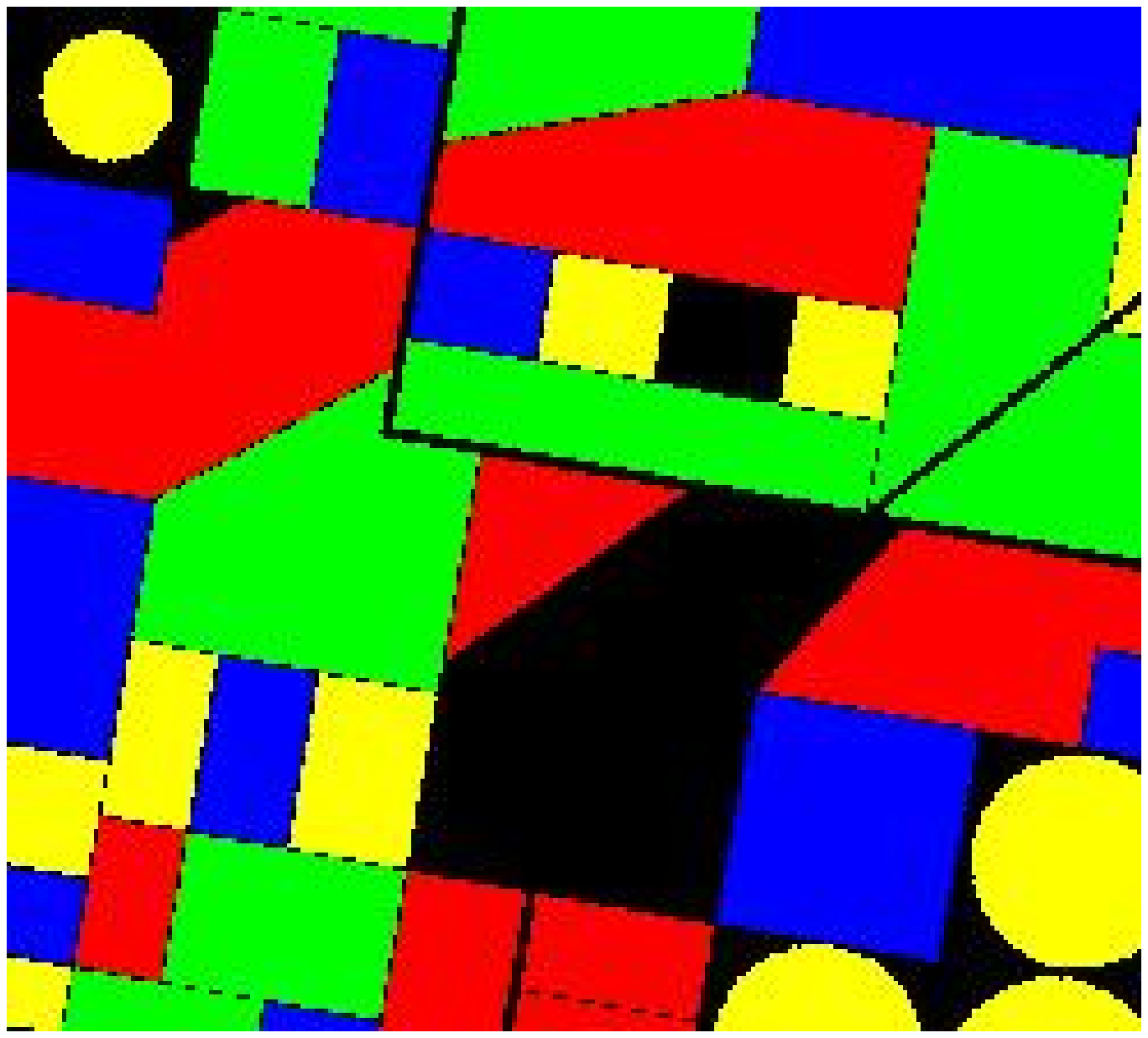}
\end{center}
\caption{\label{figure:RSimage} \it{The original (left) and the
mis-registered (right) remote sensing images for a cropland. Each
color corresponds to one land class. }}
\end{figure}

The problem of image registration is of primary importance in remote
sensing land monitoring applications which typically require the use
of a number of images acquired at different times or time sequence
data that can characterize seasonal changes or multi-annual
similarities (Defries and Townshend, 1999
\cite{DefriesTownshend1999}; Liu et al., 2006
\cite{LiuKellyGong2006}). This demands image registration and can
affect such applications as image classification, change detection,
ecological/climatological/hydrological modeling (Justice et al.,
1998 \cite{JusticeVermoteTownshend1998}; Gong and Xu, 2003
\cite{GongXu2003}) etc. Because image registration can never be
perfectly made, a mis-registration error is inevitable. It has been
suggested that mis-registration errors that are less than 0.5 pixels
are acceptable in subsequent analysis (Gong et al., 1992
\cite{GongLeDrewMiller1992}; Townshend et al. 1992
\cite{TownshendJusticeGurney1992}; Jensen, 2004 \cite{Jensen2004}).
However, this is rarely achievable and it is thus important to
assess the impact of image mis-registration.

Of a similar nature are errors due to rounding or the inaccuracy of
the measuring instruments. Besides, interference from
electromagnetic waves, clouds or other unfavorable weather
conditions can all cause errors to the remote sensing images.
Additionally, various types of human errors often factor in where a
small amount of arbitrary error maybe thrown in anywhere in the data
or any part of the data can be missing. Errors of this type are
often called gross errors, and are estimated to occur in about
$0.1\%$ to $10\%$ of the data \cite{Hampel1974}. This estimation of
the amount of errors will form the basis for our choice on the
amount of data contamination in our simulation.

We call errors discussed above broadly as data contamination. Data
contamination can cause a disastrous effect to the data quality and
may fundamentally impact subsequent analysis and inference. It is
thus of significant practical importance to answer the question:
{\it How much does data contamination impact our analysis
(classification)? Do current algorithms (classifiers) continue to
work or how much do we lose in accuracy if a remote sensing image is
mis-registered or the underlying data are contaminated?} The goal of
the present work aims to shed lights on these questions.
To gain insights into the nature of data contamination, in
particular the phenomenon of image mis-registration, it is highly
desired to approach the problem with a formal model and to give some
theoretical characterization. This forms the primary motivation of
the present work. Our focus will be on classification.

Assume the data of interest are drawn i.i.d. from some probability
distribution $G$ defined on $\mathbb{R}^p$. By treating errors as
contaminations to the probability distribution $G$, we arrive at the
following statistical model for data contamination
\begin{equation}
\label{eq:DCModel} \tilde{G}=(1-\epsilon)G+\epsilon{H}
\end{equation}
where $\tilde{G}$ is the distribution of the data after
contamination and $H$ is an arbitrary distribution.
Model~\eqref{eq:DCModel} is quite general, clearly it captures
various types of data contaminations we have discussed (not the
additive noise though). Note that, in the setting of classification,
$G$ is the joint distribution of the attributes and the label, thus
a contamination under model~\eqref{eq:DCModel} can mean that to the
attributes, or the label, or both. The $\epsilon$ in
\eqref{eq:DCModel} can be thought of as the proportion of data
(e.g., image pixels) that are ``contaminated", e.g., being flipped
in label or altered with data generated under a different
distribution $H$.

It is known that the effect of image mis-registration is determined
by resolution, scene structure and amount of registration error
(e.g., 0.5 pixels or 1 pixel, or 1.5 pixels on RMS error). In
model~\eqref{eq:DCModel}, we choose to use the proportion of pixels
that are ``contaminated" as a measure of the extent of image
mis-registration. This is to capture the essence of image
mis-registration and to uncover the relationship between the amount
of mis-registration and the resulting loss in classification
accuracy. This is different from the usual practice in the remote
sensing community where the image mis-registration is quantified in
term of a shift of a certain number of pixels. Since given the same
amount of shift, the impact on classification is highly
scene-dependent, e.g., the impact would be drastically different for
a large land consisting mainly of forests and a small land parcel
formed by corn fields and rice fields, it would then hardly be
possible to establish a generic relationship between the amount of
mis-registration and the resulting loss on the classification
accuracy.

Our contributions are as follows. We propose a statistical model for
the phenomenon of image mis-registration. This data contamination
model captures a wide range of errors such as label flipping,
measurement errors, rounding errors and accidental human errors
which occur almost ubiquitously in real applications. We study
classification under data contamination in the statistical learning
framework. A bound is obtained on the loss of classification
accuracy (term this as the data contamination bound) due to data
contamination (to the training data) in terms of its amount. This
bound allows one to give a conservative assessment on if a class of
classification algorithms, i.e., those which are universally
consistent, continue to work under data contamination.

The rest of the paper is organized as follows. In Section
\ref{section:classificationDC}, we formulate the problem of
classification under data contamination and obtain a bound on the
loss in classification accuracy in terms of the amount of data
contamination. This is followed by a discussion of related work in
statistics, remote sensing and machine learning in
Section~\ref{section:related}, and in particular we compare various
aspects of our bound with the finite sample type of bounds
established in the recently emerging area--domain adaptation. In
Section~\ref{section:simulation}, we conduct extensive simulations
on the impact of data contamination to classification performance of
SVM for a number of synthetic and real datasets under various types
of data contaminations. In Section~\ref{section:amtDC}, we briefly
discuss heuristics to estimate the amount of data contamination for
the case of image mis-registration. Finally we conclude in
Section~\ref{section:conclusion}. In this section, we also collect
results from the literature on the impact of classification
performance by AdaBoost due to label flipping; additionally, we give
insight on using data contamination as a model to understand
co-training, which is particularly useful in situations where
training data are scarce. 
\section{Classification under data contamination}
\label{section:classificationDC} Classification is an important
problem in pattern recognition. However, as discussed in
Section~\ref{section:Introduction}, especially in the context of
land-cover, land-use mapping, crop yield estimation and many other
important applications in remote sensing, the classification result
may be affected by data contamination. In this section, we will
study classification under data contamination with
model~\eqref{eq:DCModel} and derive a bound on the resulting loss in
classification accuracy. We start by an introduction of the
statistical learning framework for classification
\cite{DevroyeGyorfiLugosi1996}.
\subsection{Classification in the statistical learning framework}
In statistical learning, a classification rule (or classifier) is
defined by a map: $\mathcal{X} \rightarrow \mathcal{Y}$
where $\mathcal{X}$ is the sample space for observations and
$\mathcal{Y}$ is a finite set of labels. For simplicity, we consider
throughout a two-class problem where $\mathcal{Y}=\{0,1\}$.

Associated with each classifier, there is a performance measure
called loss function, denoted by $l(f,X,Y)$. The loss function that
is of special interest is the 0-1 loss, defined as
\begin{equation}
\label{eq:01loss} l(f,X,Y) = \left\{
\begin{array}{ll}
0   &      \mbox{if}~I_{\{f(X)>0\}}=Y \\
1   &      \mbox{otherwise}
\end{array}\right.
\end{equation}
where $f$ is a decision function and $I_{\{.\}}$ is the indicator
function. Here we call a function $f$ a decision function if a
decision rule can be written as $I_{\{f>0\}}$.

\textbf{Definition.} Let $\mathbb{P}$ be the joint probability
distribution of $X$ and $Y$. Then the risk associated with a
decision function $f$ is defined as
\begin{eqnarray}
R_{\mathbb{P}}(f) &=& \mathbb{E}_{\mathbb{P}}l(f,X,Y)=\mathbb{P}(Y
\neq I_{\{f(X)>0\}}).
\end{eqnarray}

Similarly, the empirical risk for a decision function $f$, on a
training sample $(X_1,Y_1),...,(X_n,Y_n)$, can be obtained by
replacing $\mathbb{P}$ in the above with its empirical distribution
$\hat{\mathbb{P}}_n$.

Fix a probability distribution $\mathbb{P}$ and a function class
$\mathcal{G}$, the goal of classification is to find a decision rule
$f^*_{\mathcal{G}} \in \mathcal{G}$ that minimizes
$R_{\mathbb{P}}(f)$, i.e.,
\begin{equation}
\label{eq:defLearning} f^*_{\mathcal{G}}=\arg\min_{f \in
\mathcal{G}} R_{\mathbb{P}}(f).
\end{equation}
The rule learned from the training sample $(X_1,Y_1),...,(X_n,Y_n)$,
denoted by $f_n$, can be defined similarly by substitution of
$\mathbb{P}$ with $\hat{\mathbb{P}}_n$ in \eqref{eq:defLearning}.

\textbf{Definition.} Fix a probability distribution $\mathbb{P}$.
The function that achieves the minimum risk, among all possible
decision rules, is called the Bayes rule. The corresponding risk is
called the Bayes risk and is denoted by $R_{\mathbb{P}}^*$.

For the 0-1 loss as defined in \eqref{eq:01loss} and a fixed
probability distribution, the Bayes rule is given by
\begin{equation*}
\beta(x)=I_{\{\eta(x)>0\}} 
\end{equation*}
where
\begin{equation*}
\eta(x)=\mathbb{P}(Y=1~|~X=x)-0.5
\end{equation*}
is called the Bayes decision function.

\textbf{Definition.} A classification algorithm is universally
consistent if, for all distributions $\mathbb{P}$,
\begin{equation*}
R_{\mathbb{P}}(f_n) \rightarrow_{a.s.} R_{\mathbb{P}}^*
\end{equation*}
as $n \rightarrow \infty$ where a.s. stands for almost surely.

\textbf{Notation.} To simplify notation, we adopt the following
convention. Denote $R \triangleq R_G$ and $\tilde{R} \triangleq
R_{\tilde{G}}$. Also we use $\tilde{}$ to indicate a quantity
associated with the contaminated distribution $\tilde{G}$.
In particular, $f_n$ and $\tilde{f}_n$ are the classifiers learned
from a training sample of size $n$ from $G$ and $\tilde{G}$,
respectively; and $\eta$, $\tilde{\eta}$ and $\eta^H$ are the Bayes
decision function under $G$, $\tilde{G}$ and $H$, respectively.

\subsection{A bound on the loss of classification accuracy}
In the standard setting of statistical learning theory, one is
interested in the consistency of a classifier, $f_n$, obtained via
empirical risk minimization, that is,
\begin{equation*}
R(f_n) \longrightarrow R^*
\end{equation*}
as $n \rightarrow \infty$. In such a case, the classifiers $f_n$ are
trained and tested with data generated from the same probability
distribution $G$.

In the present work, we consider a different setting where the
probability distribution, $\tilde{G}$, of the training sample
differs from that of the test sample, $G$. Of course if $G$ and
$\tilde{G}$ are ``totally" different, then there is no hope of
learning. We thus make the assumption that $G$ and $\tilde{G}$
differ by a small amount in the sense of a ``small" $\epsilon$ under
model~\eqref{eq:DCModel}. Clearly the rule learned from a training
sample under $\tilde{G}$ will be different from that under $G$.
Since the test sample is from $G$, classifier trained under
$\tilde{G}$ would typically have a larger classification error. One
important question is, how much additional classification error will
be introduced if the classifier is trained on a sample from
$\tilde{G}$ (instead of $G$) when testing on a sample generated from
$G$.

Really we wish to know how much $R(\tilde{f}_n)$ is different from
$R(f_n)$ as $n \rightarrow \infty$ for $\epsilon$ small. As we do
not have access to data from $G$, a natural proxy for $R(f_n)$ is
$R^*$ since $R(f_n) \rightarrow R^*$ as $n \rightarrow \infty$ for
consistent classifiers $f_n$. We start by the following risk
decomposition
\begin{eqnarray}
\label{eq:riskDecomp2} R(\tilde{f}_n)-R^* &=&
R(\tilde{f}_n)-R(\tilde{\eta})+R(\tilde{\eta})-R^*.
\end{eqnarray}
The $R(\tilde{\eta})-R^*$ term in \eqref{eq:riskDecomp2} can be
bounded by a term that depends only on the amount of contamination,
$\epsilon$, under some weak assumptions. This is stated as
Theorem~\ref{theorem:DCbound2}. The term
$R(\tilde{f}_n)-R(\tilde{\eta})$ can be shown to vanish as the
training sample size increases if the underlying classifier is
universally consistent. This is stated as
Theorem~\ref{theorem:consistencyDiffMeasure}. Note that here the
convergence rate may be different for different types of
classifiers.

\begin{theorem}
\label{theorem:DCbound2} If $g(x)$, the probability density function
of $G$, exists, then for data contamination with any distribution
$H$,
\begin{eqnarray*}
R(\tilde{\eta})-R^* & \leq & \frac{\epsilon}{1-\epsilon},
\end{eqnarray*}
where the equality holds if and only if the followings are true
\begin{itemize}
\item[a)]
\begin{equation*}
\epsilon=\frac{0.5-R^*}{1-R^*},
~h(x)=\frac{|\eta(x)|g(x)}{1-2R^*}, ~\mbox{and}
\end{equation*}
\item[b)] $P_{H}(Y=1|X=x)=1$ when $\eta(x)<0$, and 0 otherwise.
\end{itemize}
\end{theorem}
\textbf{Remark.}
\begin{enumerate}
\item
The bound as stated in Theorem~\ref{theorem:DCbound2} is sharp as it
is achievable under a special case as noted in the statement of the
theorem. 
\item
A related data contamination model is as follows.
\begin{equation}
\label{eq:DCModelg} d\tilde{G}(x)=
[1-\epsilon(x)]dG(x)+\epsilon(x)dH(x)
\end{equation}
such that $0 \leq \epsilon(x) \leq \epsilon<1$ for some positive
constant $\epsilon$ where $G, H, \tilde{G}$ are probability
distribution functions. Model \eqref{eq:DCModelg} allows the amount
of data contamination to be data dependent as long as the amount is
uniformly smaller than a constant. Similar result as
Theorem~\ref{theorem:DCbound2} can be obtained.
\end{enumerate}
To prepare for the proof of Theorem~\ref{theorem:DCbound2}, we have
the following lemma.
\begin{lemma}
\label{lemma:riskinSign} Let $f$ be a decision function. Further
assume $\mathbb{P}(f(X)=0)=0$. Then
\begin{eqnarray*}
R(f) & = &
0.5-\mathbb{E}\left[\eta(X).\mbox{sign}\left(f(X)\right)\right]
\end{eqnarray*}
where $sign(x)=1~\mbox{if}~ x>0~\mbox{and}~ -1~\mbox{otherwise}$.
\end{lemma}
\begin{proof}
Note that we can write
\begin{equation*}
R(f)=\mathbb{E}_{G}\left|Y-I_{\{f(X)>0\}}\right|.
\end{equation*}
Thus
\begin{eqnarray*}
R(f)
& = & \mathbb{E}\left[Y.I_{\{f(X)<0\}}\right]+\mathbb{E}\left[(1-Y).I_{\{f(X)>0\}}\right]\\
& = & 0.5+\mathbb{E}\left[(Y-0.5).I_{\{f(X)<0\}}\right] +\mathbb{E}\left[(0.5-Y).I_{\{f(X)>0\}}\right]\\
& = &
0.5-\mathbb{E}\left[\eta(X).\mbox{sign}\left(f(X)\right)\right].
\end{eqnarray*}
\end{proof}
\noindent The posterior probability $\tilde{\eta}(x)+0.5$ under the
contaminated distribution $\tilde{G}$ can be written as
\begin{eqnarray*}
&&\tilde{\eta}(x)+0.5  =
[1-\alpha_{\epsilon}(x)](\eta(x)+0.5)+\alpha_{\epsilon}(x)(\eta^{H}(x)+0.5)
\end{eqnarray*}
where
\begin{eqnarray*}
\alpha_{\epsilon}(x) &=& \epsilon h(x)[(1-\epsilon)g(x)+\epsilon
h(x)]^{-1}.
\end{eqnarray*}
Here $g$ and $h$ are the continuous density or discrete probability
functions corresponding to $G$ and $H$, respectively. Then
\begin{eqnarray*}
\tilde{\eta} & = &
(1-\alpha_{\epsilon})\eta+\alpha_{\epsilon}\eta^H.
\end{eqnarray*}

\begin{proof}[Proof of Theorem~\ref{theorem:DCbound2}]
By Lemma~\ref{lemma:riskinSign}, we have
\begin{eqnarray*}
R(\tilde{\eta})-R^*
&=& \mathbb{E}\left[\eta(sign(\eta)-sign(\tilde{\eta}))\right] =
2\mathbb{E}\left[|\eta|.I_{\{\eta\tilde{\eta}<0\}}\right].
\end{eqnarray*}
Next notice that, if
\begin{eqnarray*}
\eta\tilde{\eta} & = &
\alpha_{\epsilon}\eta^{2}\left[\frac{(1-\alpha_{\epsilon})}{\alpha_{\epsilon}}+\frac{2\eta^H}{2\eta}
\right]<0,
\end{eqnarray*}
then this implies
\begin{eqnarray*}
2|\eta| & \leq &
\frac{\alpha_{\epsilon}}{(1-\alpha_{\epsilon})}=\frac{\epsilon}{1-\epsilon}\frac{h(x)}{g(x)}.
\end{eqnarray*}
Hence,
\begin{eqnarray}
R(\tilde{\eta})-R^* & \leq &
2\mathbb{E}\left[|\eta|.I_{\left\{2|\eta|\leq
\frac{\epsilon}{1-\epsilon}\frac{h(X)}{g(X)}
\right\}}\right]\label{eq:L1}\\
& \leq & \frac{\epsilon}{1-\epsilon}\mathbb{E}\frac{h(X)}{g(X)}\label{eq:L2}\\
& = & \frac{\epsilon}{1-\epsilon}.\nonumber
\end{eqnarray}

The equality in \eqref{eq:L1} holds if and only if
$\eta^H=-\frac{1}{2}$, or, $P_{H}(Y=1|X=x)=1$ when $\eta(x)<0$, and
0 otherwise, i. e. for the same observation $X=x$, the worst rule
under $H$ assigns a completely oppositive class membership w.r.t.
that under $G$. Further, the equality in \eqref{eq:L2} holds if and
only if
\begin{equation*}
2|\eta(x)|  =  \frac{\epsilon}{1-\epsilon}\frac{h(x)}{g(x)},
\end{equation*}
which implies
\begin{eqnarray*}
2\mathbb{E}|\eta| & = & \frac{\epsilon}{1-\epsilon}
\end{eqnarray*}
since $\int h(x)dx=1$. Thus,
\begin{equation*}
\epsilon=\frac{2\mathbb{E}|\eta|}{1+2\mathbb{E}|\eta|}=\frac{0.5-R^*}{1-R^*}
\end{equation*}
by Lemma~\ref{lemma:riskinSign}. This concludes the proof.
\end{proof}
\begin{theorem}
\label{theorem:consistencyDiffMeasure} Suppose a classification
algorithm is universally consistent. Then, under data contamination
model~\eqref{eq:DCModel}, we have
\begin{equation*}
R(\tilde{f}_n) \rightarrow R(\tilde{\eta})
\end{equation*}
as $n \rightarrow \infty$.
\end{theorem}
The proof of Theorem~\ref{theorem:consistencyDiffMeasure} relies on
the following lemma.
\begin{lemma}
\label{lemma:ruleConvergence} Assume
$\mathbb{P}\left(\eta(X)=0\right)=0$. If $R(f_n) \rightarrow R^*$,
then the decision induced by $f_n$ converges to the Bayes rule in
probability as $n \rightarrow \infty$.
\end{lemma}
\textbf{Remark.} Theorem 2 of Bartlett and Tewari
\cite{BartlettAmbuj2007} implies that the decision rule given by SVM
converges to the Bayes rule. Lemma~\ref{lemma:ruleConvergence} is
more general in that it applies to all consistent rules. 
\begin{proof}
Without loss of generality, assume the decision function $f_n$ is
already centered, i.e., the corresponding decision rule can be
written as $I_{\{f_n>0\}}$. From Lemma~\ref{lemma:riskinSign}, we
have
\begin{equation*}
R(f_n)=0.5-\mathbb{E}\left(\eta(X)*\mbox{sign}(f_n(X))\right).
\end{equation*}
Let
\begin{equation*}
\xi_n(x)=|\mbox{sign}(\eta(x))-\mbox{sign}(f_n(x))|,
\end{equation*}
then $\xi_n(x)$ takes two values $\{0,2\}$. We have
\begin{eqnarray*}
&&R(f_n)-R^* \\
&=&\mathbb{E}(\eta(X))\left[\mbox{sign}(\eta(x))-\mbox{sign}(f_n(x))\right]
\\
&=& \mathbb{E}|\eta_m(X)|.\xi_n(X).
\end{eqnarray*}
Thus, $\mathbb{P}(\xi_n(X)=2) \rightarrow 0$ by assumption $R(f_n)
\rightarrow R^*$ as $n \rightarrow \infty$. That is,
$I_{\{f_n(X)>0\}}$ converges to $I_{\{\eta(X)>0\}}$ in probability
as $n \rightarrow \infty$.
\end{proof}

\noindent\begin{proof} [Proof of
Theorem~\ref{theorem:consistencyDiffMeasure}] By universal
consistency and Lemma~\ref{lemma:ruleConvergence}, we have
\begin{equation*}
\int I_{\left\{\mbox{sign}(\tilde{f}_n(X)) \neq
\mbox{sign}(\tilde{\eta}(X)) \right\}}d\tilde{G}(x) \rightarrow
0.
\end{equation*}
Thus
\begin{equation*}
\int I_{\left\{\mbox{sign}(\tilde{f}_n) \neq
\mbox{sign}(\tilde{\eta}) \right\}}dG \rightarrow 0,
\end{equation*}
implying that, as $n \rightarrow \infty$,
\begin{equation*}
R(\tilde{f}_n) \rightarrow R(\tilde{\eta}).
\end{equation*}
\end{proof}

By risk decomposition \eqref{eq:riskDecomp2} as well as
Theorem~\ref{theorem:DCbound2} and
Theorem~\ref{theorem:consistencyDiffMeasure}, we arrive at a sharp
asymptotic data contamination bound as
\begin{equation}
\label{eq:dataContBound2}
\frac{\epsilon}{1-\epsilon}+O\left(c(n)\right).
\end{equation}
where $c(n)=R(\tilde{f}_n) - R(\tilde{\eta})$ indicates the rate of
convergence with $c(n) \rightarrow 0$ as $n \rightarrow \infty$. 

Bound~\eqref{eq:dataContBound2} implies that, when the amount of
data contamination is ``small", i.e., $\epsilon \rightarrow 0$, we
can make
\begin{equation*}
|R(\tilde{f}_n)-R^*| \rightarrow 0.
\end{equation*}
That is, as long as a classifier is consistent in the standard
setting and the amount of contamination is small in the sense of a
small $\epsilon$, this classifier suffers very little from data
contamination. This explains why, empirically, classifiers such as
SVM or others work well even when a small fraction of labels are
randomly flipped.

Theorem~\ref{theorem:consistencyDiffMeasure} relies on the universal
consistency of a classifier. Fortunately, several of the currently
most popular classifiers are universally consistent, for example,
SVM \cite{Steinwart2002} and Adaboost with early stopping
\cite{BartlettTraskin2007}.
\section{Related work}
\label{section:related} The study of data analysis and statistical
inference under data contamination has been a long-standing research
topic in statistics and machine learning. The earliest work can be
traced back to at least a half century ago, see, for example, Tukey
\cite{Tukey1960} for a survey on sampling from contaminated
distribution. Extensive studies have been carried out since under
the name of robust estimation (\cite{Huber1972, Hampel1974}),
measurement error model (\cite{Fuller1987, carrollDelaige2009,
DelaigeFan2009}) etc. However, work along this line concerns
primarily problems on regression or estimation.

Relevant literature in remote sensing, however, has been sparse.
Swain el al \cite{SwainVanderbilt1982} investigated the impact of
image mis-registration to classification. However, this work is
purely empirical and their results depend highly on the underlying
scenes in the image; for example, even under the same amount of
mis-registration, the impact would be considerably different on
images formed primarily by large forest lands and those formed by
many small patches of different land types such as corns and plants.
Additionally, Townshend el al \cite{TownshendJusticeGurney1992}
considered the impact of image mis-registration to change detection.
Xu et al \cite{XuDickson2009} study parameter estimation for a
simple linear model under measurement errors due to a mismatch of
locations and scales.

Related machine learning literature is much richer. Such work can be
broadly divided into two stages. The first stage, roughly before
year 2005, mostly deals with data contamination in the form of label
flipping and empirical study of its impact on the performance of
various classifiers. This includes Dietterich \cite{Dietterich1998}
and Breiman \cite{RF} which evaluate the robustness of learning
algorithms such as bagging, AdaBoost and Random Forests against
label flipping. Other work includes
(\cite{Quinlan1986,Sloan1988,ZhuWu2004}) and references therein. The
second or the current stage, which is closely related to the present
work, deals with domain adaptation. Domain adaptation is a broader
concept than data contamination in that it does not specify
explicitly the nature of the difference between the source (or
training) distribution and the target (or test) distribution as long
as their difference is small whereas data contamination almost
exclusively refers to model~\eqref{eq:DCModel}. There have been
numerous papers published on domain adaptation, including
applications, theory and methods, and it is beyond the scope of the
present paper to give a detailed account here. Work that is closest
to ours include
(\cite{Ben-DavidBCKPV2010,MansourMohriRostamizadeh2009}) (see also
references therein). In particular, Ben-David et al
\cite{Ben-DavidBCKPV2010} established the following bound.

\begin{theorem}[\cite{Ben-DavidBCKPV2010}]
\label{theorem:Ben-DavidBCKPV2010} Let $\mathcal{H}$ be a hypothesis
space of VC dimension $d$. If $\mathcal{U}_S$, $\mathcal{U}_T$ are
unlabeled samples of size $m'$ each, drawn from the source
distribution $\mathcal{D}_S$ and the target distribution
$\mathcal{D}_T$ respectively, then for any $\delta \in (0,1)$, with
probability at least $1-\delta$ (over the choice of the samples),
for every $h \in \mathcal{H}$, the difference between the error
rates $\epsilon_S$ and $\epsilon_T$ satisfies
\begin{eqnarray*}
&&\epsilon_T(h) - \epsilon_S(h) \nonumber\\
&\leq& \frac{1}{2}\hat{d}_{\mathcal{H} \Delta \mathcal{H}}
(\mathcal{U}_S,
\mathcal{U}_T)+4\sqrt{\frac{2d\log(2m')+\log(2/\delta)}{m'}}+\lambda
\label{eq:BenDavidBound2010}
\end{eqnarray*}
where $\lambda$ is defined by
\begin{equation*}
\lambda=\arg \min_{h \in \mathcal{H}} [\epsilon_S(h)+\epsilon_T(h)]
\end{equation*}
with the subscripts $S,T$ indicating quantities related to the
source and target, respectively.
\end{theorem}

The bound established in \cite{MansourMohriRostamizadeh2009} is
similar in nature which replaces the VC dimension in
\cite{Ben-DavidBCKPV2010} with the Rademacher complexity
\cite{BartlettMendelson2001}. However, there are important
differences between the bound in
Theorem~\ref{theorem:Ben-DavidBCKPV2010} or that in
\cite{MansourMohriRostamizadeh2009} and ours (i.e.,
Theorem~\ref{theorem:DCbound2}).

\begin{itemize}
\item[(1)] The nature of bounds is different. The bounds in
(\cite{Ben-DavidBCKPV2010,MansourMohriRostamizadeh2009}) are finite
sample learning generalization type of bounds while our bound is a
large sample bound (i.e., asymptotic bound).
\item[(2)]
The quality of the bounds is different. The bounds in
\cite{Ben-DavidBCKPV2010} are union bounds that rely on the
Vapnik-Chervonekis (VC) dimension \cite{Vapnik1998}, and are often
quite loose (\cite{MansourMohriRostamizadeh2009} uses the Rademacher
complexity \cite{BartlettMendelson2001} but still quite loose). In
contrast, our bound is a sharp bound asymptotically. Assume the
underlying function class has a finite VC dimension and let $m'
\rightarrow \infty$, then the bound in
Theorem~\ref{theorem:Ben-DavidBCKPV2010} becomes $\epsilon+\lambda$,
which is looser than our bound $\epsilon/(1-\epsilon) \approx
\epsilon$ for small $\epsilon$. Since the $\lambda$ term depends on
the difficulty of the underlying problem and generally does not
vanish, in no way would the bounds in \cite{Ben-DavidBCKPV2010}
imply ours.

To better appreciate the difference in the quality of the bounds
when the sample size increases, we will show an example where the
data is generated by a two-component Gaussian mixture and
contaminated by Cauchy data (See Section~\ref{section:simulation}
for details on the Gaussian mixture and the Cauchy). Since it is not
easy to directly compute $\lambda$, we replace it with its lower
bound $\arg \min_{h \in \mathcal{H}} \epsilon_S(h) + \arg \min_{h
\in \mathcal{H}} \epsilon_T(h)$, which are estimated as the error
rates of SVM on the data when the training sample size is large.
Figure~\ref{figure:cmpBounds} shows the asymptotic data
contamination bounds of ours and that established in
\cite{Ben-DavidBCKPV2010} for the amount of data contamination
varying from $\{0.01,0.02,0.03,0.04,0.05,0.10\}$. One can see that
here the Ben-David et al bound \cite{Ben-DavidBCKPV2010} is much
looser than ours, and for this particular Gaussian mixture data, the
Ben-David et al bound is not very informative as it quickly
approaches $0.5$.
\begin{figure}[h]
\centering
\begin{center}
\hspace{0cm}
\includegraphics*[scale=0.32,clip]{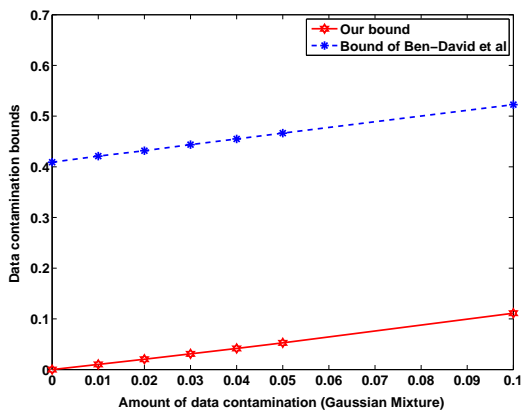}
\end{center}
\caption{\label{figure:cmpBounds} \it{Comparison of data
contamination bound for Gaussian mixture data with $\epsilon \in
\{0.01,0.02,0.03,0.04,0.05,0.10\}$.}}
\end{figure}
\item[(3)] Whereas our bound applies only to
universally consistent classifiers, the bound in
\cite{Ben-DavidBCKPV2010} applies only to classifiers from a
function class with a finite VC dimension. This is a limitation that
cannot be overlooked. For example, the function class corresponding
to the Gaussian kernel (see discussion after example 1.9 in
\cite{Steinwart2005} and the fact that the Gaussian kernel is a
universal kernel), or the polynomial kernel (if no upper bound is
imposed on the degree of polynomials), or the one nearest neighbor
classifier, or the tree-based classifier (without regularization)
all have an infinite VC dimension. Consequently, the bound in
(\cite{Ben-DavidBCKPV2010}) excludes some of the best classifiers
available today, including SVM with the Gaussian kernel, Boosting
(or Bagging \cite{Bagging}) on tree-based classifiers etc while our
bound clearly does not have such a restriction.
\end{itemize}

\section{Experiments}
\label{section:simulation} Empirical studies are performed on three
different types of datasets, $3$ synthetic datasets, $10$ UC Irvine
datasets \cite{UCI} and a simulated remote sensing image. For each
dataset, four different types of data contaminations are applied to
the training set and classification accuracy evaluated on the
uncontaminated test set. SVM is used as the underlying classifier
due to its universal consistency \cite{Steinwart2002} and the
availability of a widely used software implementation (libsvm
\cite{LIBSVM}).

The five different types of data contaminations are as follows.
\begin{itemize}
\item[$C_0$.]
Randomly flip the labels of a randomly selected subset of
observations from a fixed class.
\item[$C_1$.]
Randomly flip the labels of a randomly selected subset of
observations from all classes.
\item[$C_2$.]
Randomly select a subset of observations and replace the feature
values of each with that of a randomly chosen observation (the
labels are kept). Call this feature swapping.
\item[$C_c$.]
Replace a randomly selected subset of observations with Cauchy data
with the labels kept.
\item[$C_g$.]
Replace a randomly selected subset of observations with Gaussian
data with the labels kept.
\end{itemize}
\begin{figure}[ht]
\centering
\begin{center}
\hspace{0cm}
\includegraphics*[scale=0.4,clip]{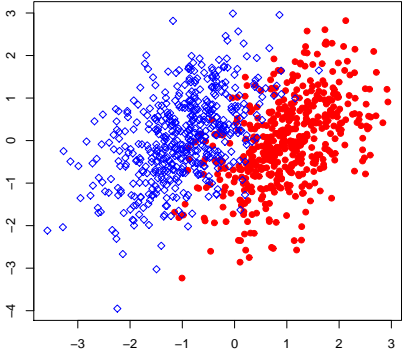}
\end{center}
\caption{\label{figure:plotGaussian2D} \it{Scatter plot of $1000$
observations generated i.i.d. from Gaussian
$\mathcal{N}(\mu,\Sigma)$ with $\mu=(1,0)$ and $\Sigma=A^tA$ with
entries of $A$ generated i.i.d. from $\mathcal{U}[0,1]$. Data from
the two classes are represented as diamonds and solid circles,
respectively. }}
\end{figure}
\begin{figure*}[htp]
\centering
\begin{center}
\hspace{0cm}
\includegraphics*[scale=0.28,clip]{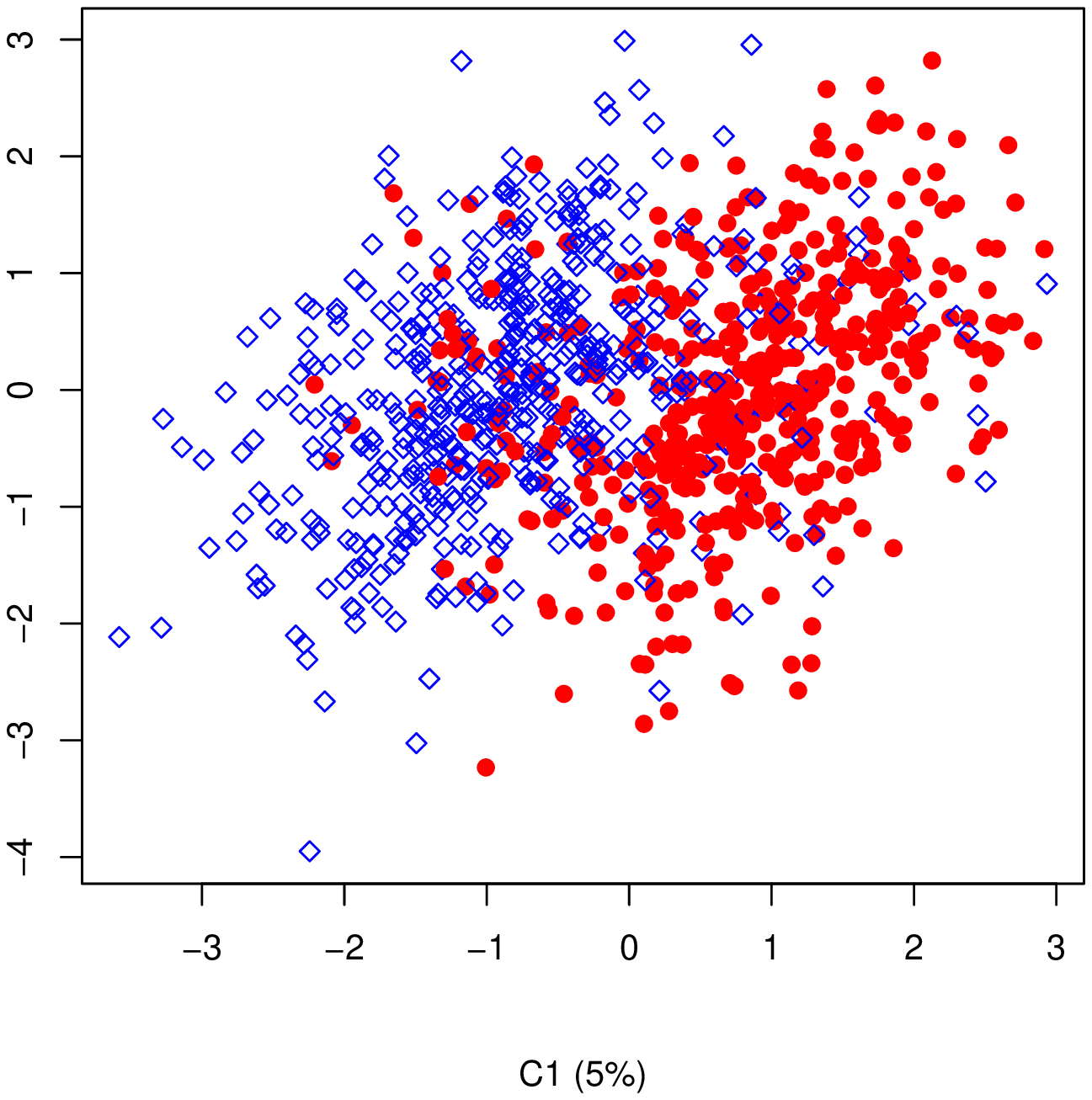}
\includegraphics*[scale=0.28,clip]{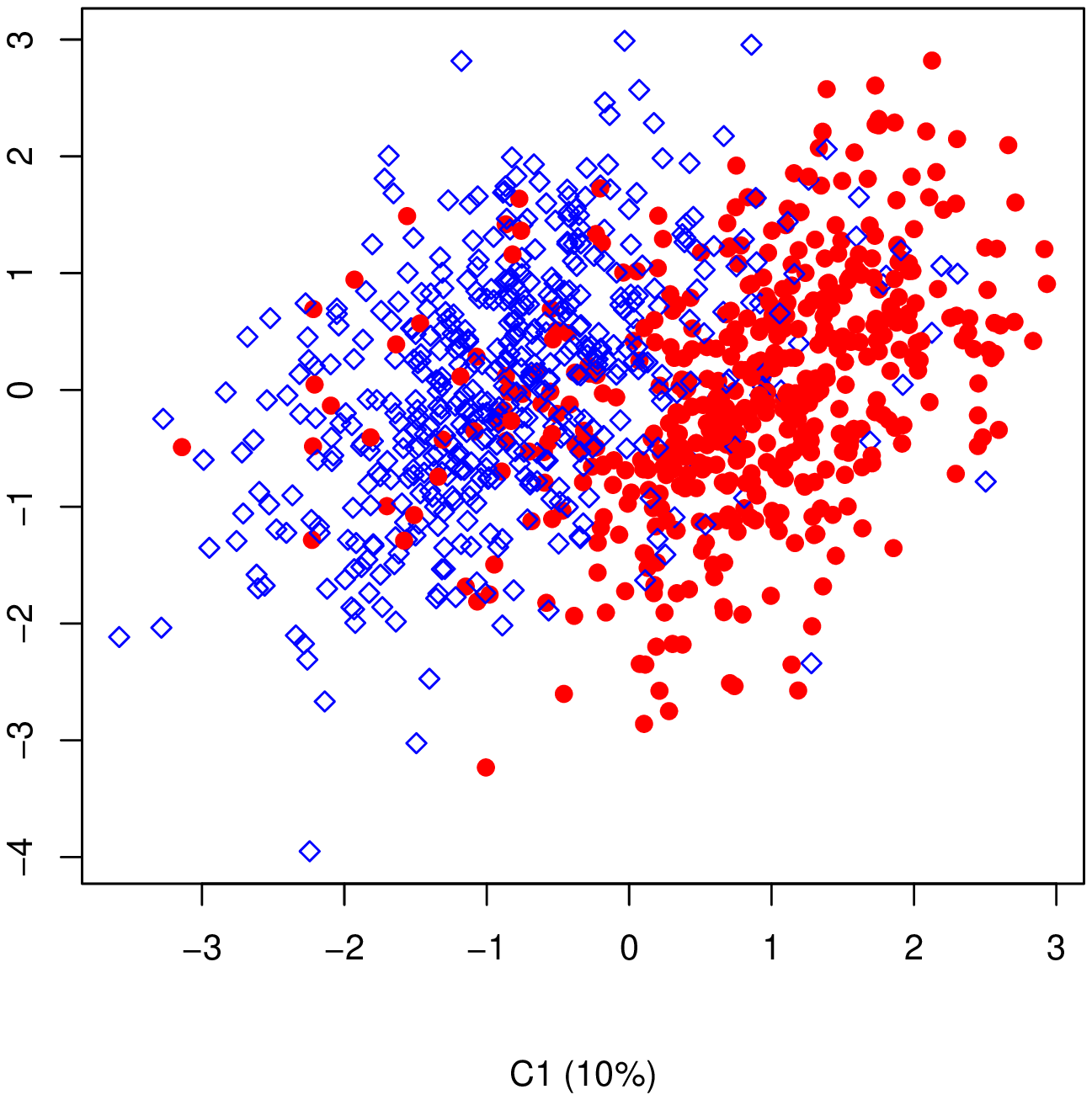}
\end{center}
\begin{center}
\hspace{0cm}
\includegraphics*[scale=0.28,clip]{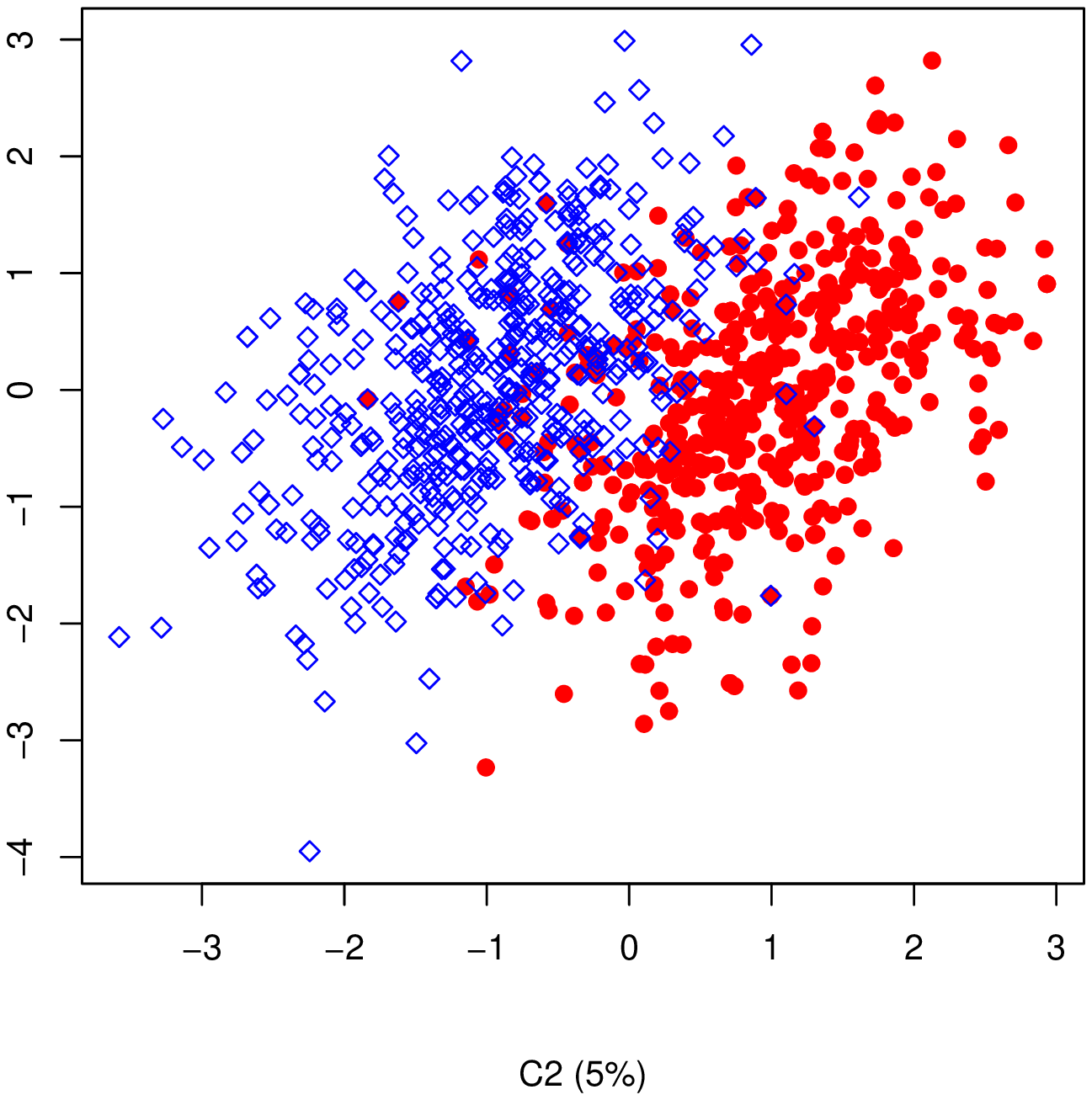}
\includegraphics*[scale=0.28,clip]{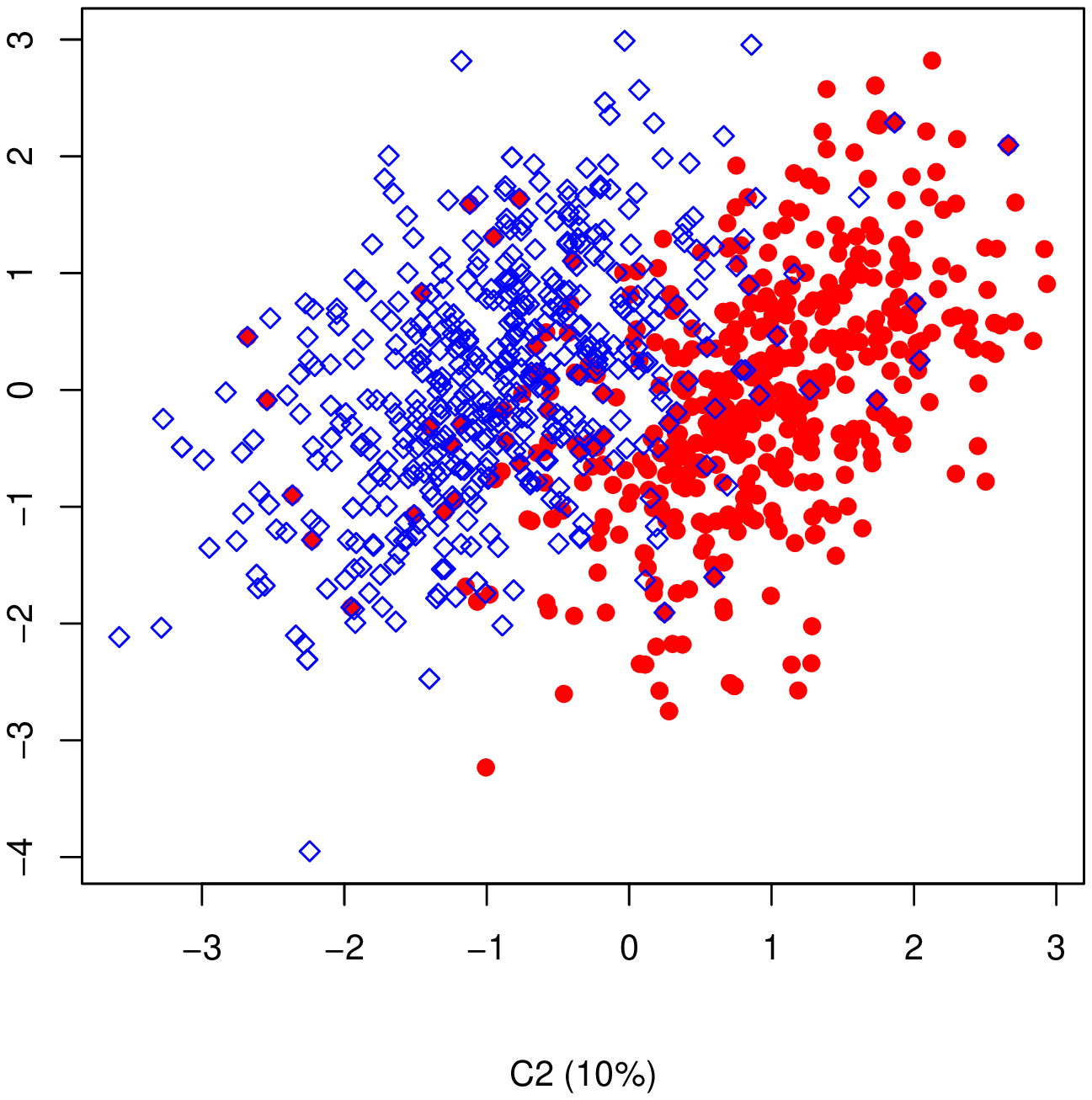}
\end{center}
\begin{center}
\hspace{0cm}
\includegraphics*[scale=0.28,clip]{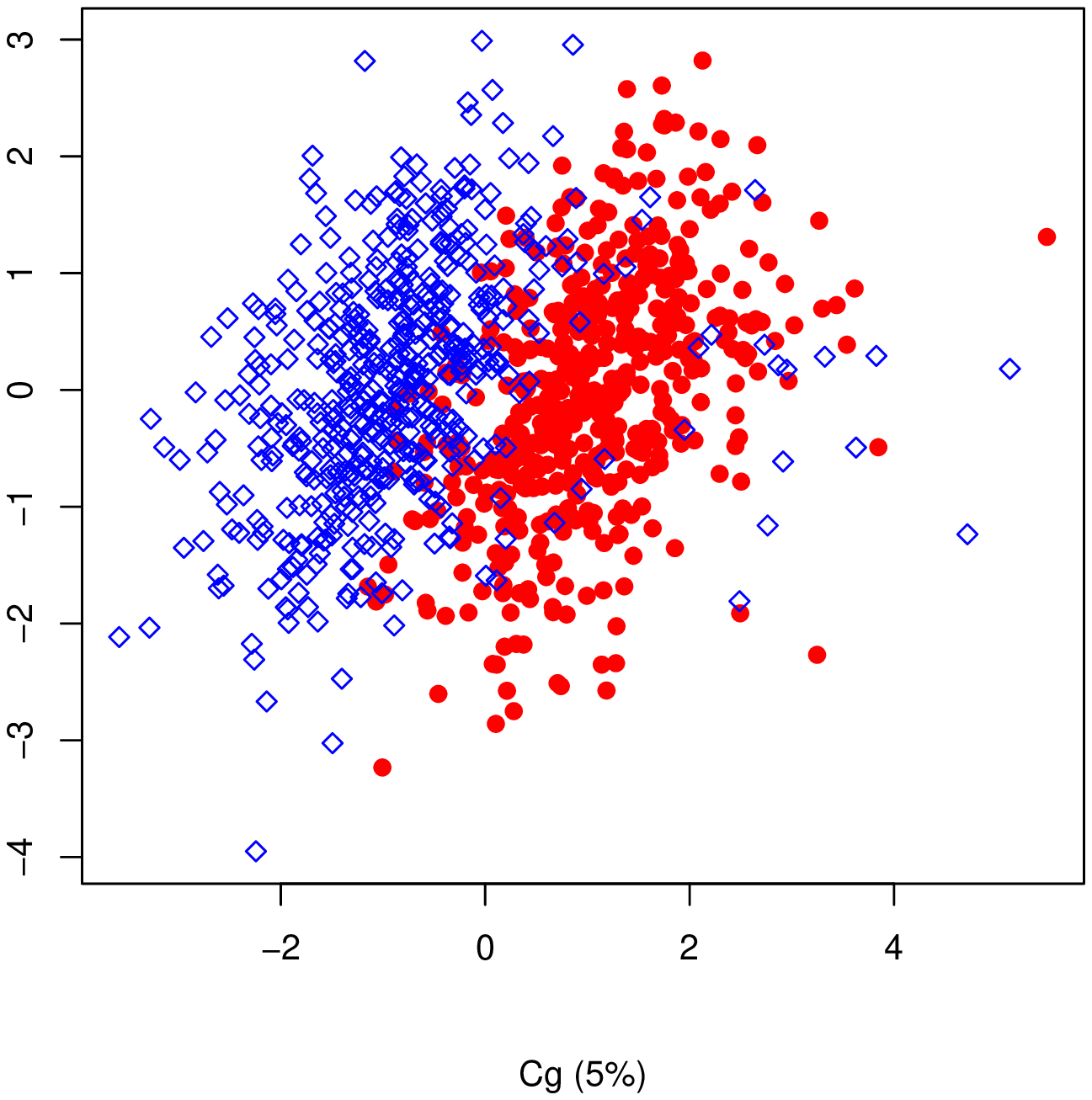}
\includegraphics*[scale=0.28,clip]{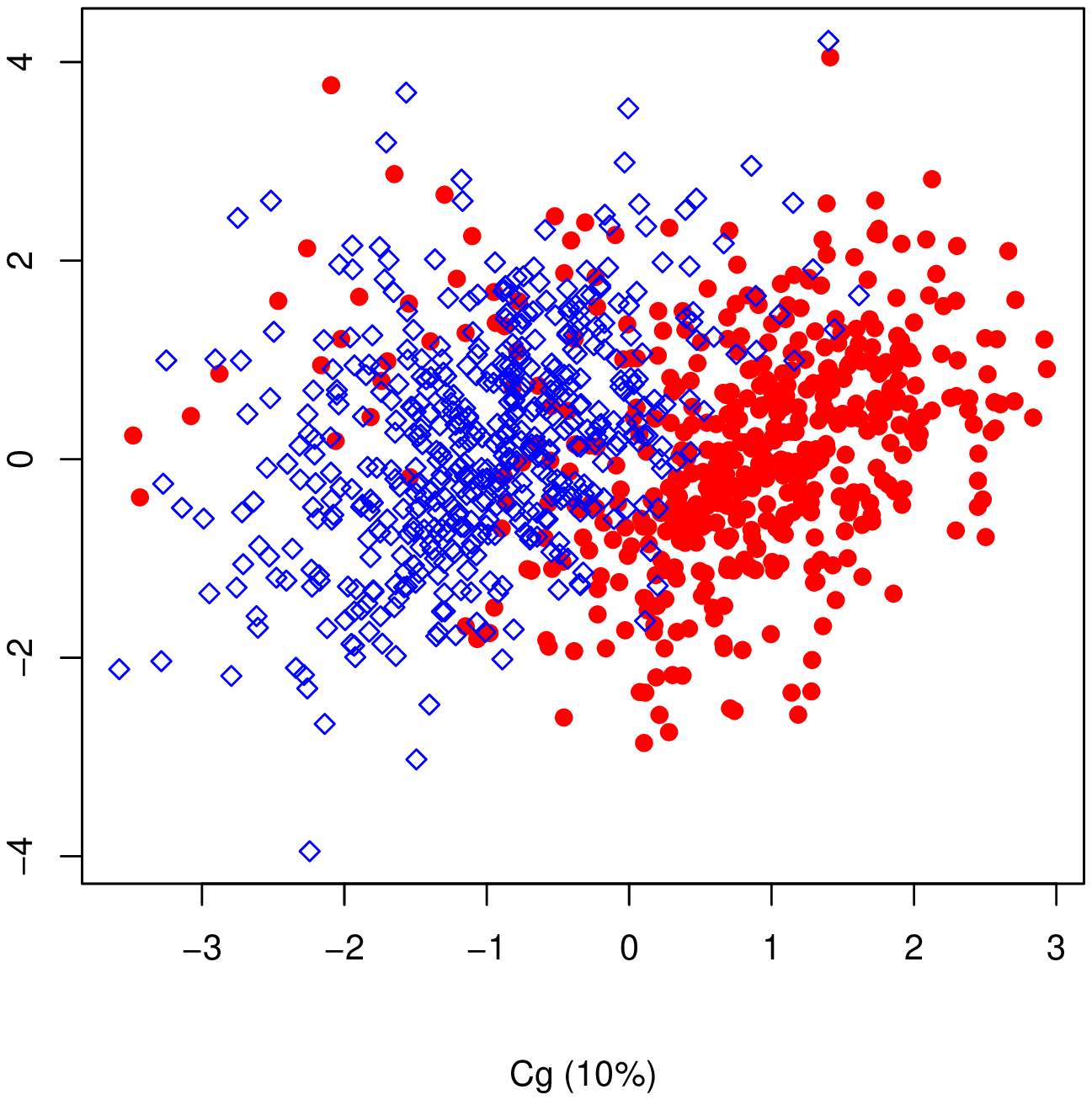}
\end{center}
\begin{center}
\hspace{0cm}
\includegraphics*[scale=0.28,clip]{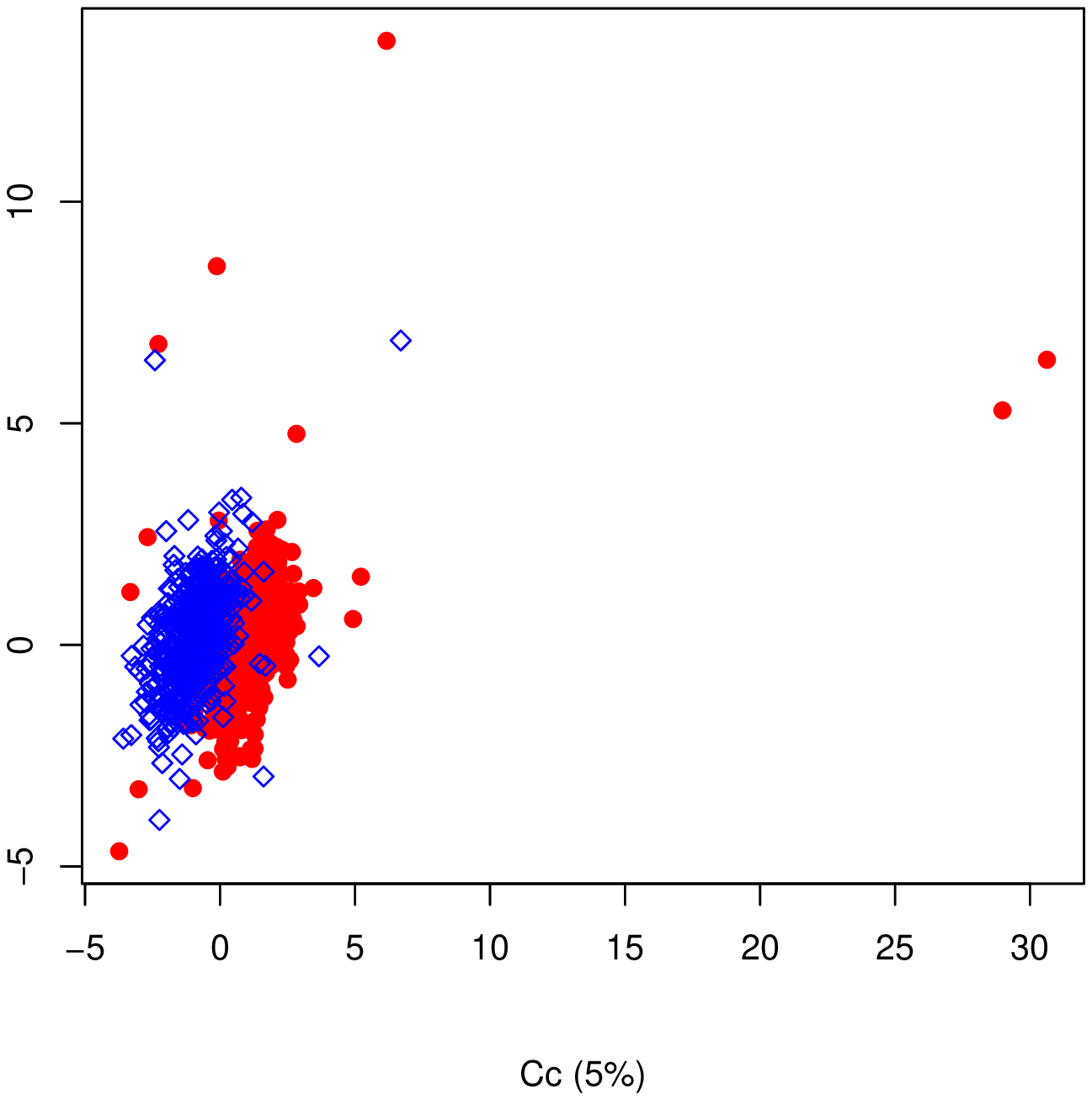}
\includegraphics*[scale=0.28,clip]{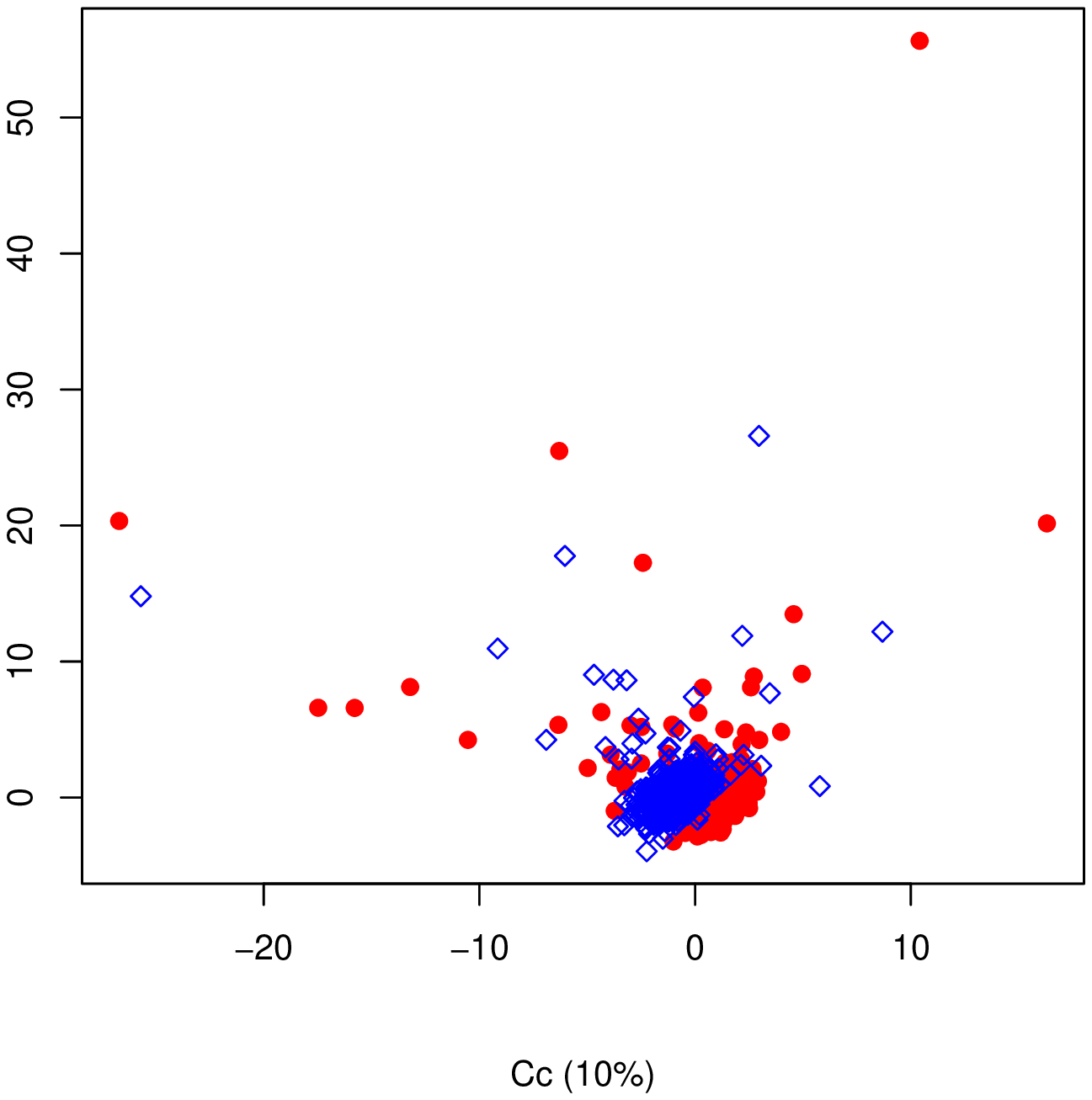}
\end{center}
\caption{ \it{Illustration of the effect of different types of data
contamination. The original Gaussian data are displayed in
Figure~\ref{figure:plotGaussian2D}. The $4$ rows of plots correspond
to $C_1,C_2,C_g, C_c$, respectively and figures in the left and
right columns are for data contamination at $5\%$ and $10\%$,
respectively. Data from the two classes are represented as diamonds
and solid circles, respectively. }} \label{figure:plotGaussianDC}
\end{figure*}
\noindent $C_0, C_1,...,C_g$ are used to simulate data contamination
of different natures.
\begin{itemize}
\item
$C_1$ and $C_2$ are expressly designed to simulate image
mis-registration, which we believe capture important aspects of
image mis-registration.
\item
$C_c$ and $C_g$ are used to simulate gross errors. $C_g$ is for
errors with a Gaussian nature while $C_c$ is for errors with a heavy
tail, that is, the error could be very large and this is to simulate
accidental human error, for example, a shift in decimal place of a
number.
\item
Additionally, we also attempt to simulate extremely large errors by
scaling the centers of the Gaussian and Cauchy by a factor of $100$,
that is, the centers are multiplied by $100$ coordinate-wisely.
These are denoted by $C_{g100}$ and $C_{c100}$, respectively.
\item
$C_0$ is used to simulate a class of unfavorable situations where
data contamination occurs in part of the data space. Such cases
typically make classification more challenging. In contrast, other
simulations are more or less average cases as the data contamination
occurs uniformly across the whole data space.
\end{itemize}
For $C_g$, the replacement Gaussian data is generated i.i.d. from
$\mathcal{N}(\mu,\Sigma)$ with $\mu$ and $\Sigma$ calculated
empirically on the non-contaminated training set. For $C_c$, the
Cauchy data is generated i.i.d. according to
\begin{equation*}
Z/W, \quad\mbox{for}~Z \sim \mathcal{N}(\mu,\Sigma), ~W \sim
[\Gamma(0.5,2)]^{1/2}
\end{equation*}
with $Z$ and $W$ independent where $\Gamma(0.5,2)$ is a random
variable generated from a Gamma distribution with parameters $0.5$
and $2$. For each run, $\mu$ is generated uniformly from the
interval $[\min(X),\max(X)]$ and $\Sigma$ estimated empirically from
the training set.

For an illustration of the effect of these different types of data
contamination, see Figure \ref{figure:plotGaussian2D} for the
original data and Figure~\ref{figure:plotGaussianDC} for the data
after contamination of different types.
\subsection{Synthetic data}
The three synthetic datasets used in our experiment are the Gaussian
mixture data, the four-class and the nested-square data. The
Gaussian mixture data are used to simulate cases with a linear
decision boundary while the four-class and the nested-square
datasets are for cases where the decision boundary is highly
nonlinear and non-convex. For each of the $3$ datasets, we take
$80\%$ for training and the rest for test. Then $100$ instances of
data contamination are applied and loss in classification accuracy
are averaged. This is repeated and results are averaged. The
Gaussian kernel is used with SVM for all three synthetic datasets.

\begin{figure}[ht]
\centering
\begin{center}
\hspace{0cm}
\includegraphics*[scale=0.34,clip]{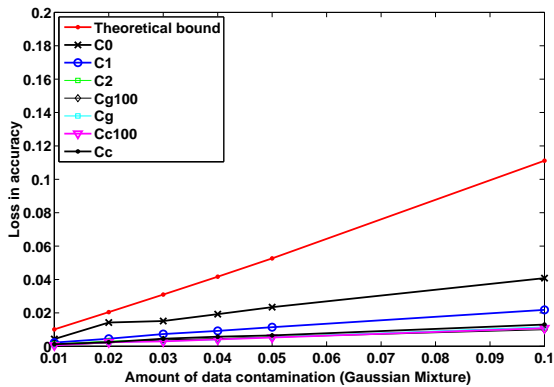}
\end{center}
\caption{\label{figure:diffGaussian} \it{Empirical and theoretical
data contamination bound for data generated from a Gaussian mixture
with $\epsilon \in \{0.01,0.02,0.03,0.04,0.05,0.10\}$.}}
\end{figure}

The Gaussian mixture data are generated according to the following
\begin{equation*}
\Delta\mathcal{N}(\mu,\Sigma_{10 \times
10})+(1-\Delta)\mathcal{N}(-\mu,\Sigma_{10 \times 10})
\end{equation*}
with $\mathbb{P}(\Delta=1)=\mathbb{P}(\Delta=0)=\frac{1}{2}$ and
$\Sigma_{10 \times 10}=A^T A$ for entries of $A$ generated i.i.d.
uniform from $[0, 1]$, with $\mu=(0.5,...,0.5)^T$. Data points with
$\Delta=1$ are assigned label $1$ and those with $\Delta=0$ are
assigned label $2$. The sample size for the training set and test
set are $1000$ and $2000$, respectively.
Loss in classification accuracy under data contamination of
different types and at different amounts are shown in
Figure~\ref{figure:diffGaussian}. Note that here we are using only
the first term in \eqref{eq:dataContBound2} as an estimate of the
overall loss in classification accuracy while ignoring the second
term, thus when the training sample size is not large enough, some
adjustment (in the order of $O(c(n))$) might be required.

\begin{figure}[ht]
\centering
\begin{center}
\includegraphics*[scale=0.24,clip]{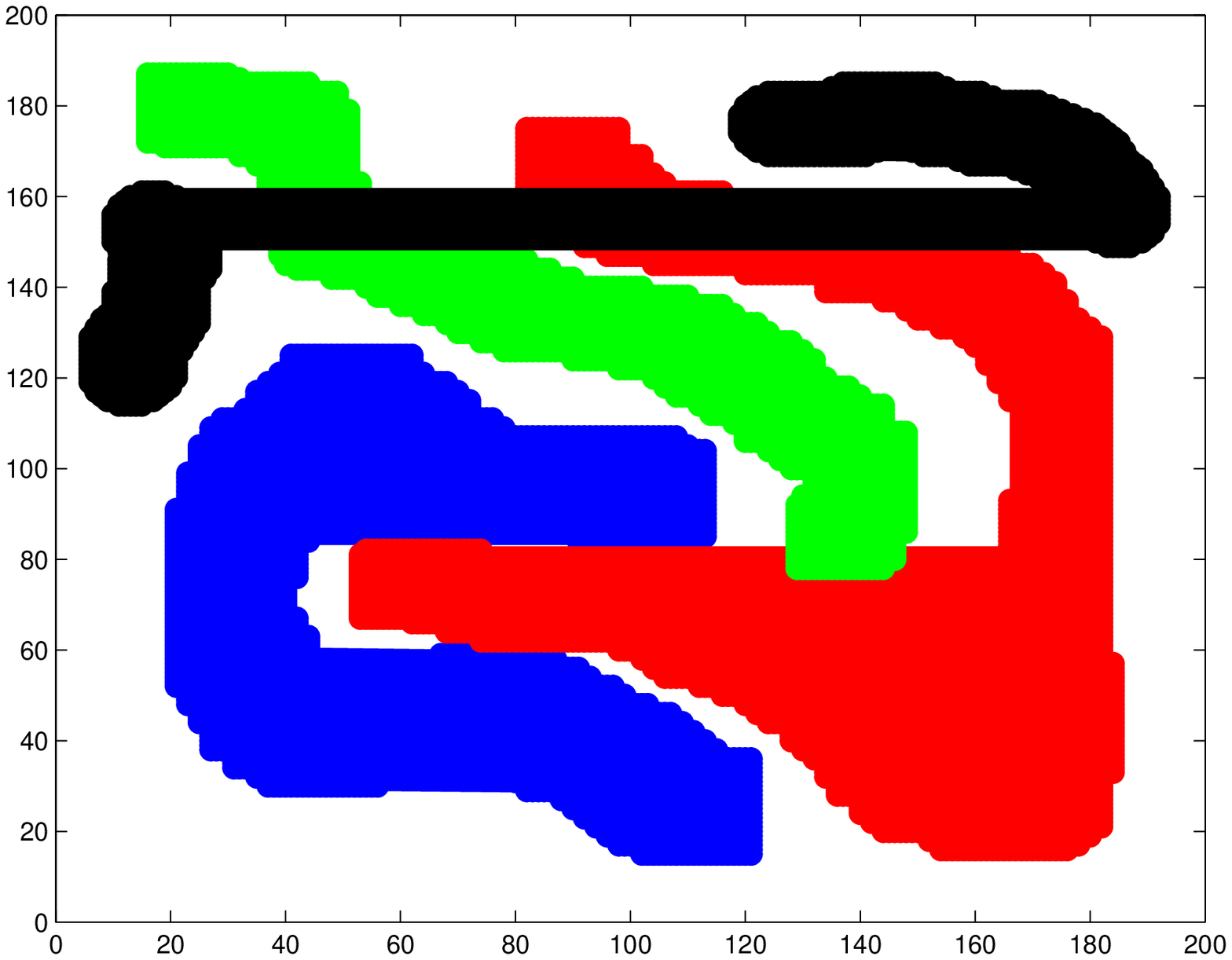}
\quad
\includegraphics*[scale=0.24,clip]{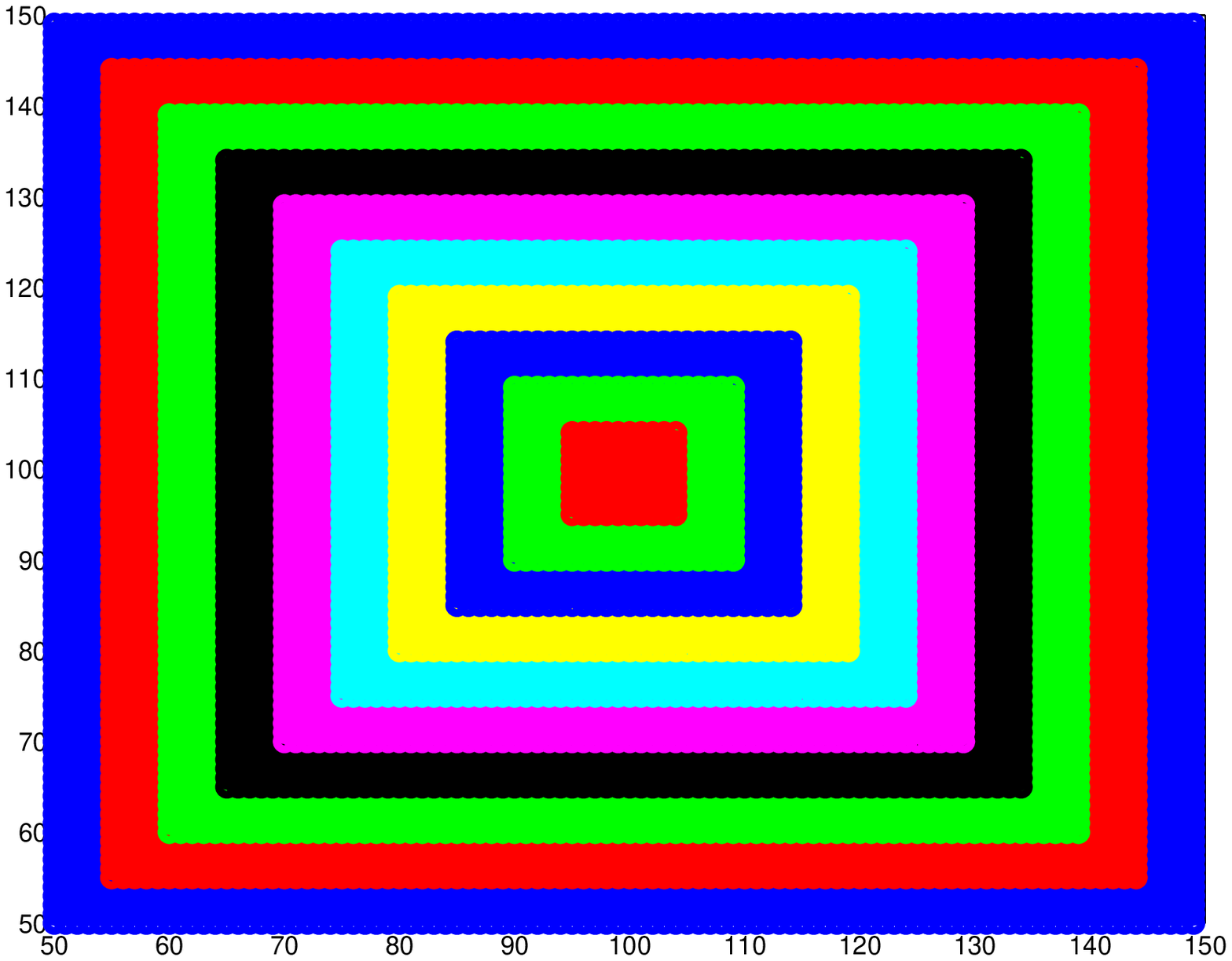}
\end{center}
\caption{\label{figure:THdata} \it{The four-class and nested-square
data. Different colors correspond to points from different
classes.}}
\end{figure}

\begin{figure}[h]
\centering
\begin{center}
\hspace{0cm}
\includegraphics*[scale=0.24,clip]{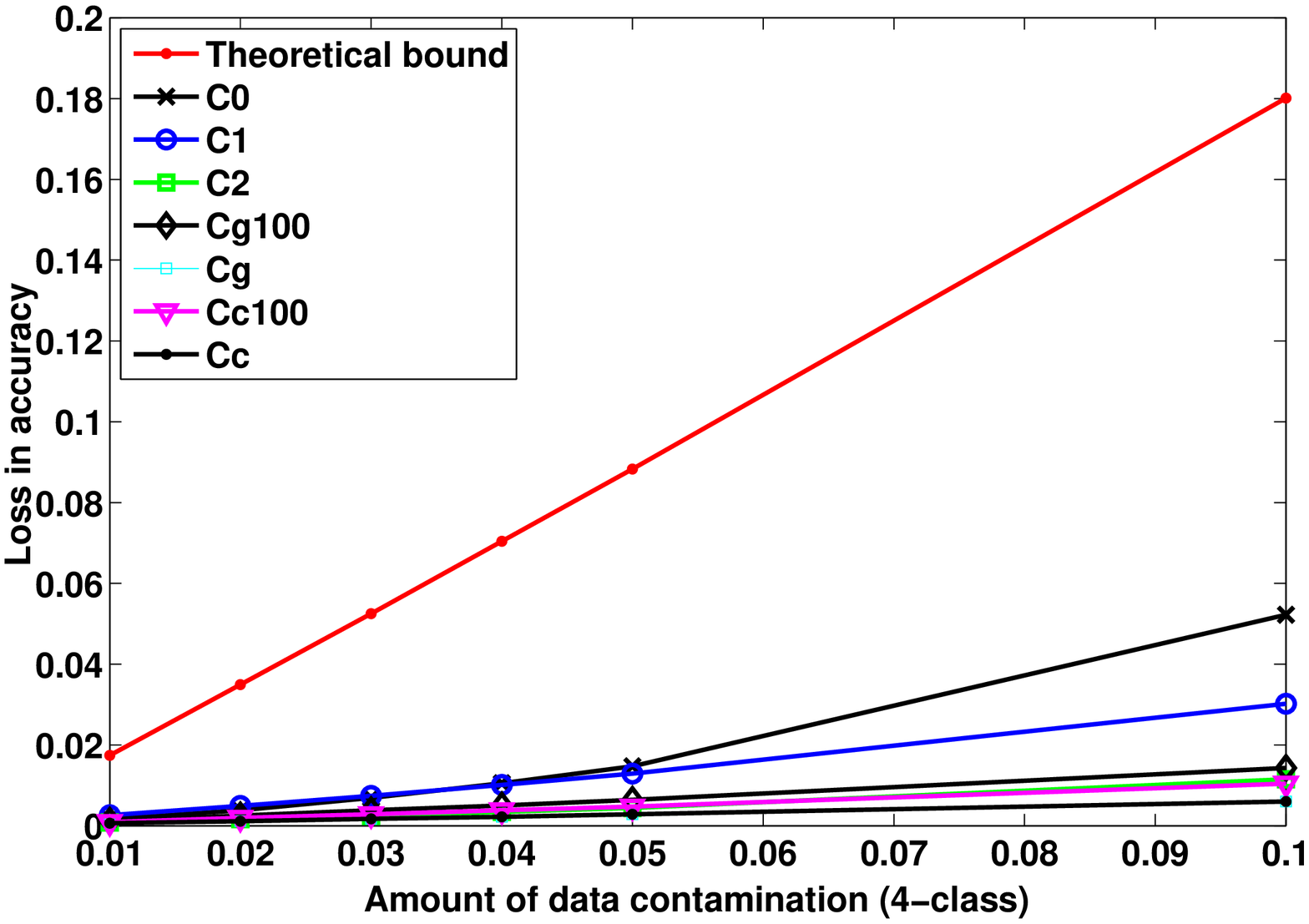}
\includegraphics*[scale=0.24,clip]{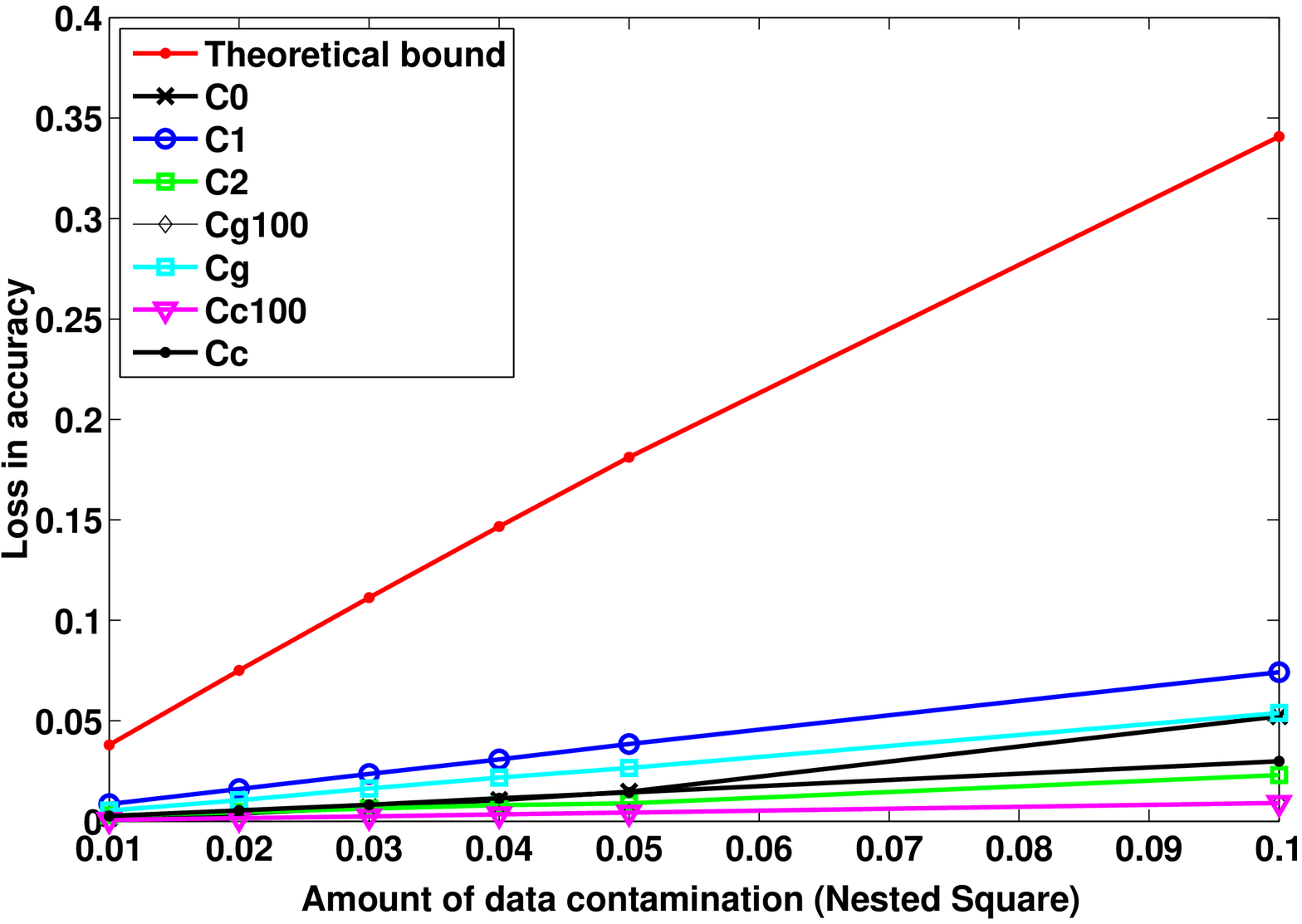}
\end{center}
\caption{\label{figure:diffBayesS1} \it{Empirical and theoretical
data contamination bound for the 4-class and the nested-square
datasets with $\epsilon \in \{0.01,0.02,0.03,0.04,0.05,0.10\}$.}}
\end{figure}

The four-class and nested-square datasets were originally used to
demonstrate the superior performance of a class of projectable
classifiers for data with a highly complex decision boundary
\cite{HoKleinberg1996}. Figure~\ref{figure:THdata} is a plot of
these two datasets and the data contamination bounds are shown in
Figure~\ref{figure:diffBayesS1}. Note that the bound as established
in \eqref{eq:dataContBound2} is for 2-class classification. When
there are multiple classes, we can get a bound by repeatedly apply
the the 2-class bound. Let the class distribution be denoted by
$\{w_1,...,w_J\}$ such that $w_1 \geq ... \geq w_J$. Then we get the
following multi-class bound
\begin{equation*}
\label{eq:mclassBound}
\frac{\epsilon}{1-\epsilon}\left\{1+(w_2+...+w_J)\alpha+...+(w_{\{J-1\}}+w_J)\alpha^{J-2}\right\}
\end{equation*}
where $\alpha=1-\frac{\epsilon}{1-\epsilon}$ and $\epsilon$ is the
mount of contamination. This is used as the theoretical bound in our
simulations when there are more than two classes. 
\subsection{UC Irvine datasets}
A total of $10$ datasets are taken from the UC Irvine Machine
Learning Repository \cite{UCI} in our experiment. A summary of these
datasets is provided in Table~\ref{table:UCIdatasets} and more
details can be found from \cite{UCI}.
\begin{table}[htp]
\caption{\it{Summary of the UC Irvine datasets used in our
experiment. }}
\begin{center}
\begin{tabular}{|c||c|c|c|c|}
\hline
              &Training     &Testing         &Features        &Classes\\
\hline
$imageSeg$     &210               &2100              &19                &7\\
\hline
$Vowel$        &528              &462               &10                &11\\
\hline
$Satellite ~images$  &4435       &2000              &36                &6\\
\hline
$Glass$       &214               &--                &10                &6\\
\hline
$Vehicle$     &946               &--                &18                &4\\
\hline
$German ~credit$         &1000    &--                &24                &2\\
\hline
$Yeast$       &1484               &--               &8                 &10\\
\hline
$Wine~quality$ &1599             &--               &11                &6\\
\hline
$Musk$        &6598              &--                &168               &2\\
\hline
$Magic ~gamma$ &19020             &--                &10                &2\\
\hline
\end{tabular}
\end{center}
\label{table:UCIdatasets}
\end{table}

Some data sets come with predetermined training and test sets, which
includes the image segmentation, vowel and satellite image datasets.
Otherwise we split the data into a training and test set. For small
to medium sized datasets, i.e., Glass, Vehicle, German Credit, Yeast
and Wine Quality (red wine), we take $80\%$ of the data for training
and the rest for test. For large datasets, i.e., the Musk and Magic
Gamma Telescope, $20\%$ and $10\%$, respectively, of the data are
set aside for training and the rest for test. For each dataset,
$100$ instances of data contamination are applied to the training
set and the resulting data contamination bounds are averaged. This
is repeated
and results averaged.

\begin{figure*}[htp]
\centering
\begin{center}
\hspace{0cm}
\includegraphics*[scale=0.24,clip]{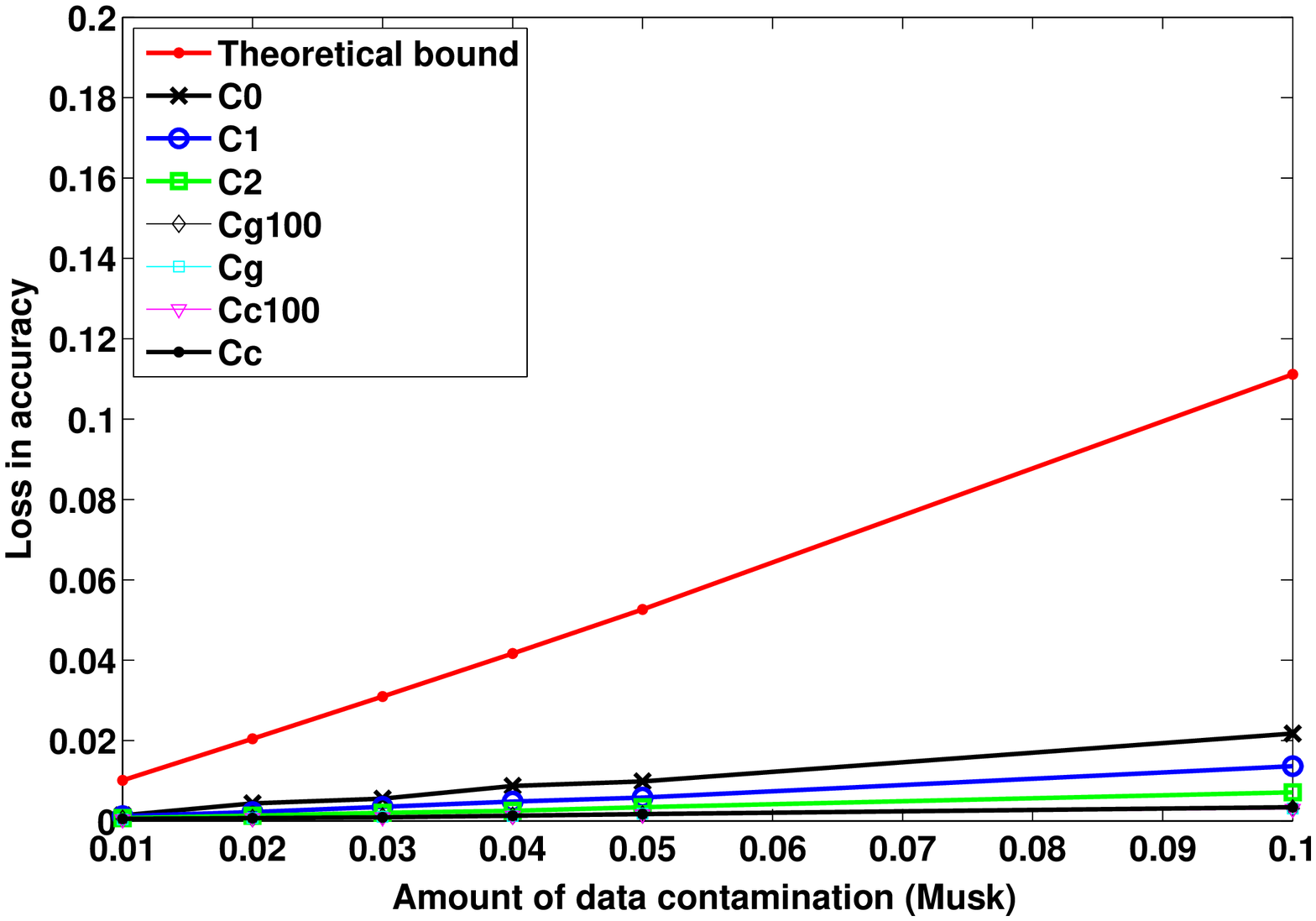}
\includegraphics*[scale=0.24,clip]{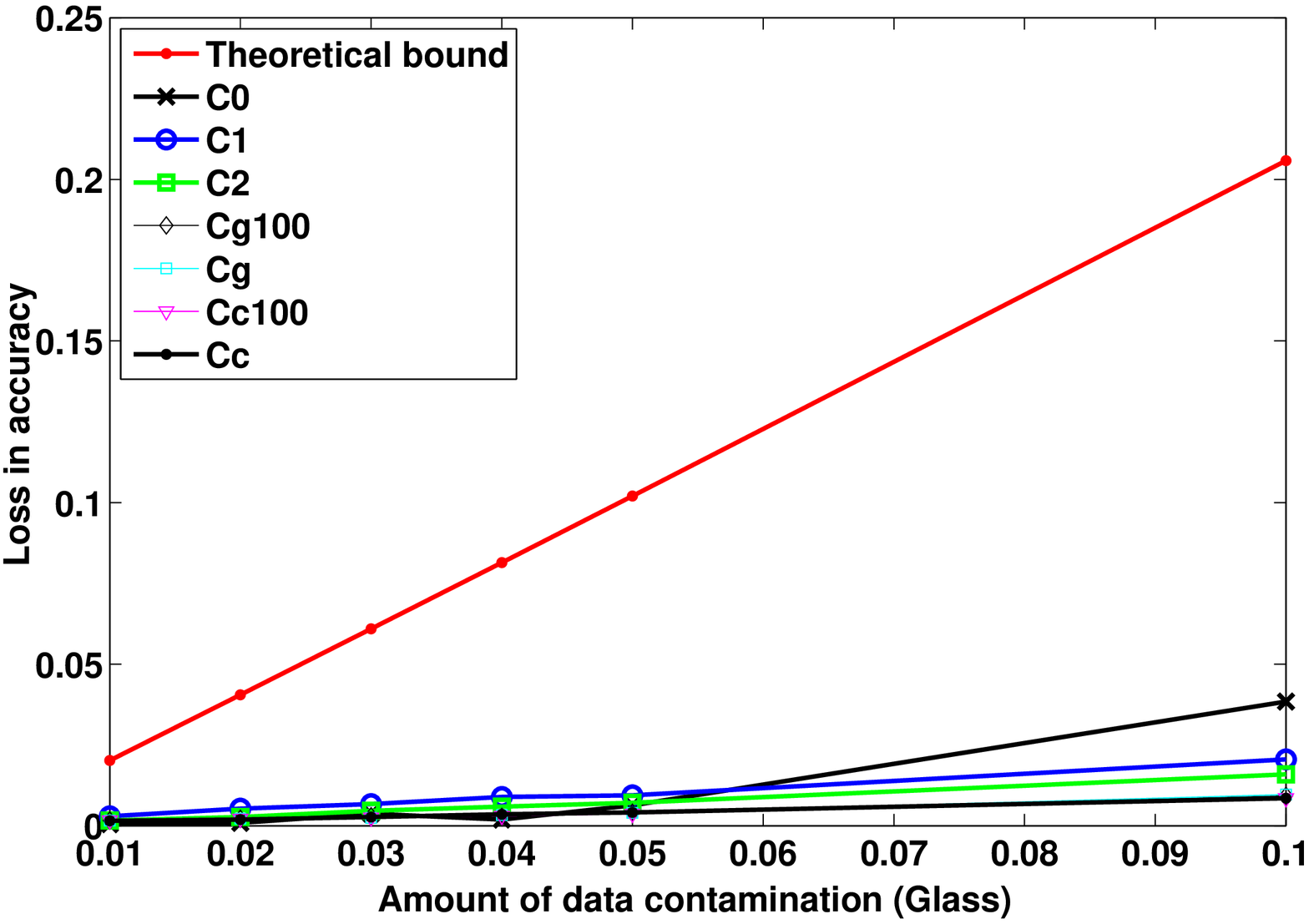}
\includegraphics*[scale=0.24,clip]{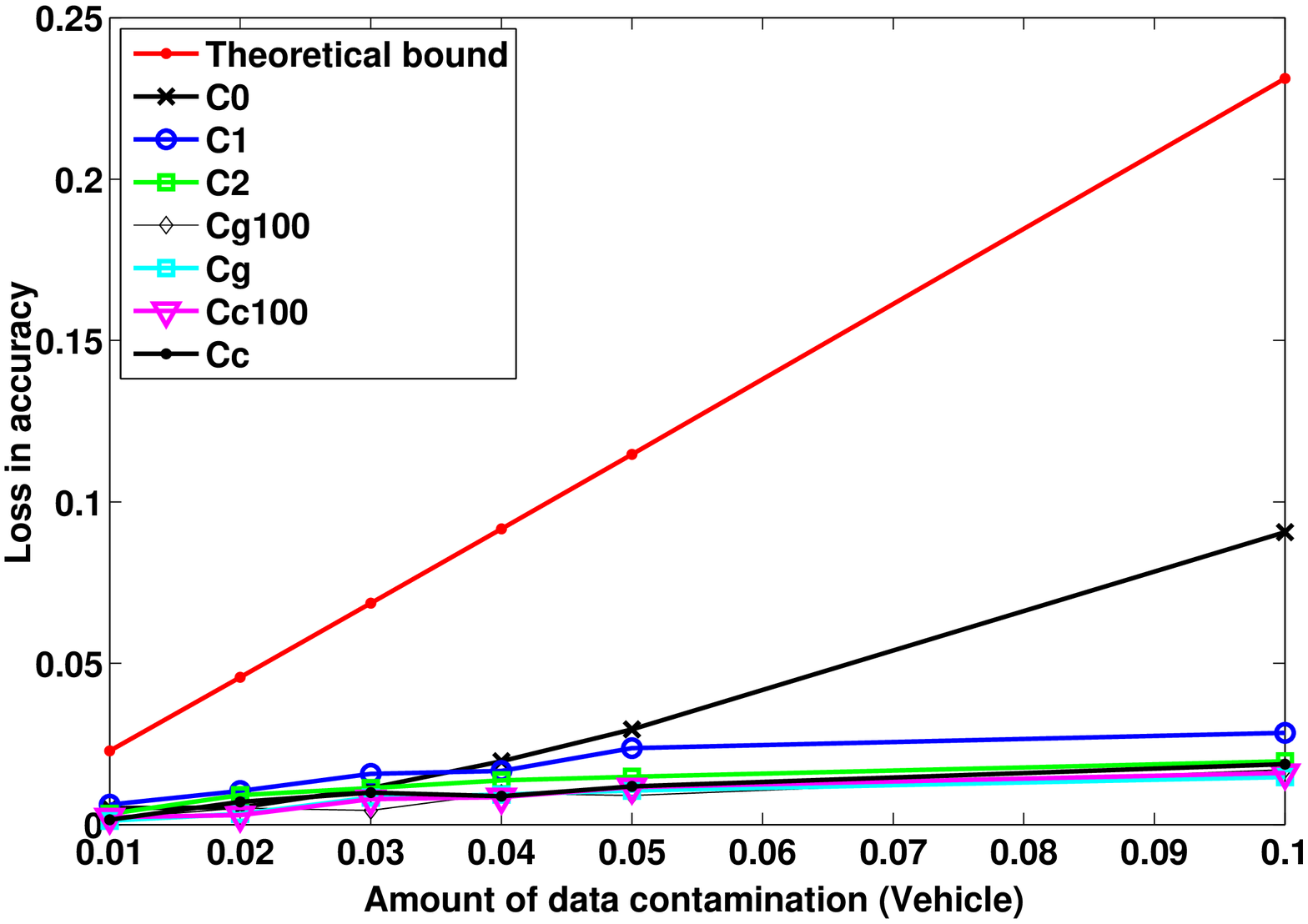}
\includegraphics*[scale=0.24,clip]{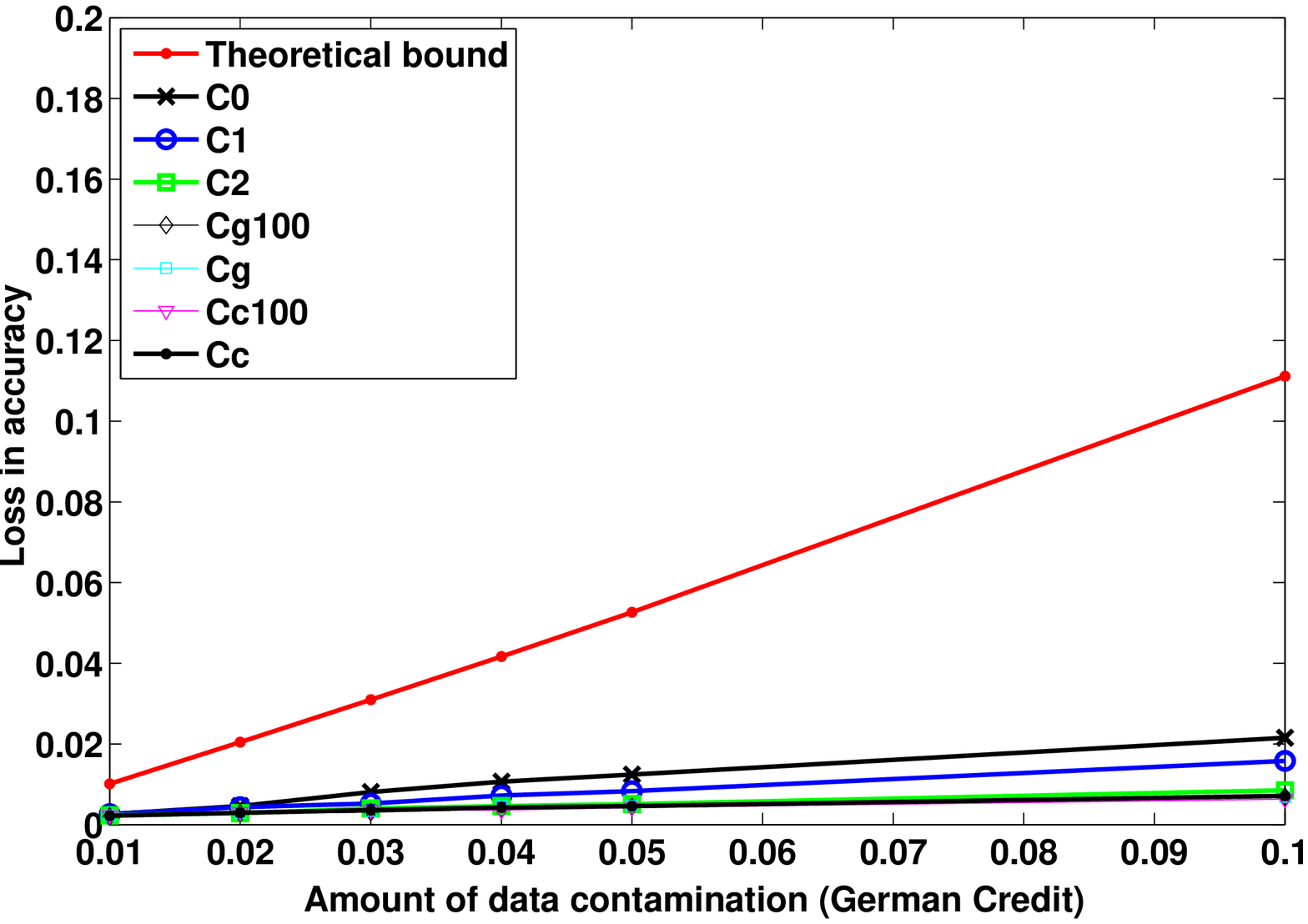}
\includegraphics*[scale=0.24,clip]{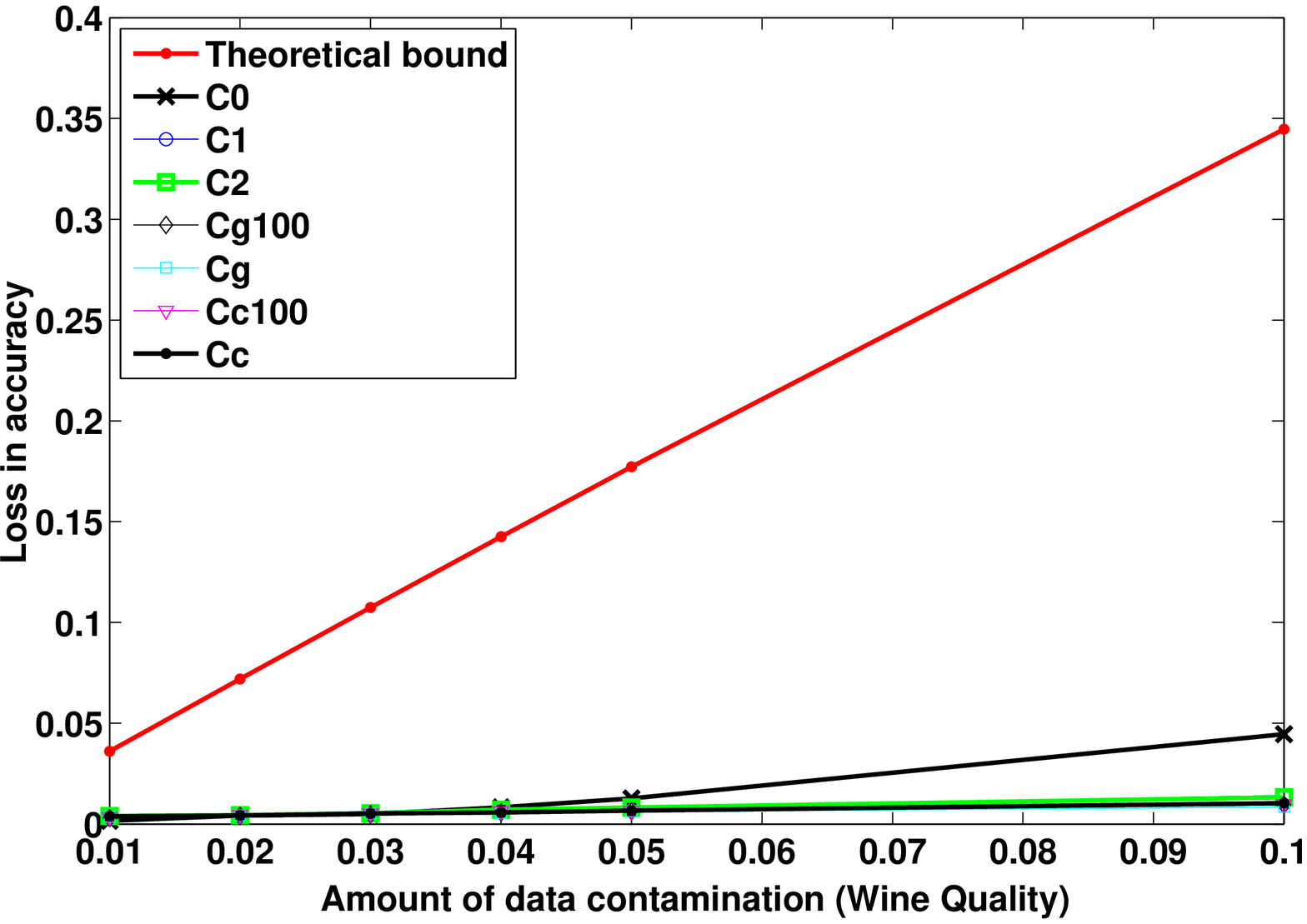}
\includegraphics*[scale=0.24,clip]{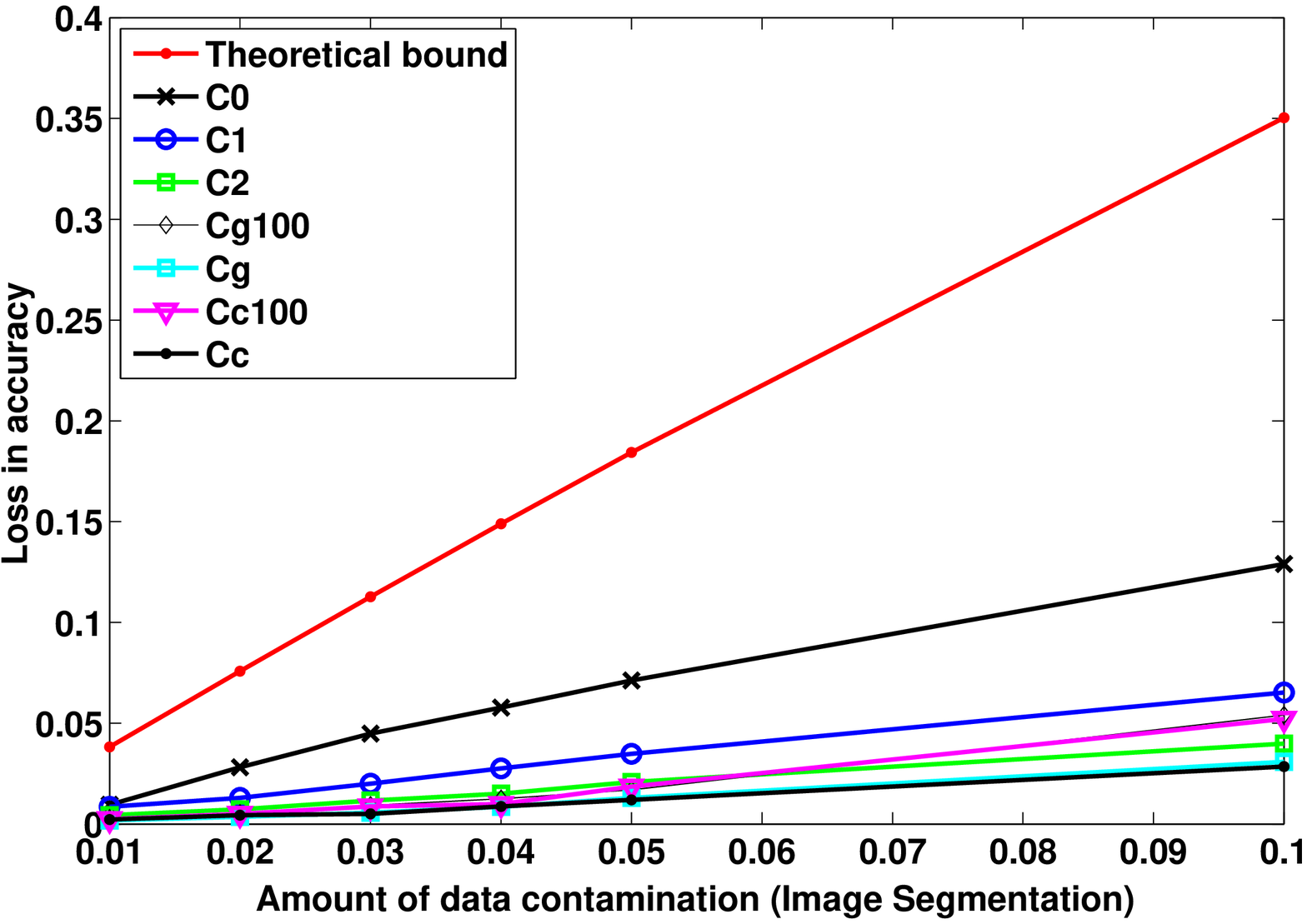}
\end{center}
\caption{\label{figure:diffBayesS2} \it{Empirical and theoretical
data contamination bound for UC Irvine datasets (only $6$ of them
are shown here so that they can be placed in the same page, the rest
are similar) with $\epsilon \in
\{0.01,0.02,0.03,0.04,0.05,0.10\}$.}}
\end{figure*}

The Gaussian kernel is used for all except the image segmentation
dataset where a polynomial kernel with degree $3$ is used. Tuning
parameters for SVM are chosen so that the classification performance
matches that reported in the literature (see, for example,
references cited in the description of each dataset in \cite{UCI}).
Some datasets are linearly scaled to $[0,1]$ so as to speed up the
painfully slow optimization of the SVM package; this includes the
Musk, Magic Gamma, Satellite image, Vehicle, and the Wine quality
dataset. The data contamination bounds by SVM on the UC Irvine
datasets are plotted in Figure~\ref{figure:diffBayesS2}. 
\subsection{Remote sensing image}
\begin{figure}[ht]
\centering
\begin{center}
\hspace{0cm}
\includegraphics*[scale=0.45,clip]{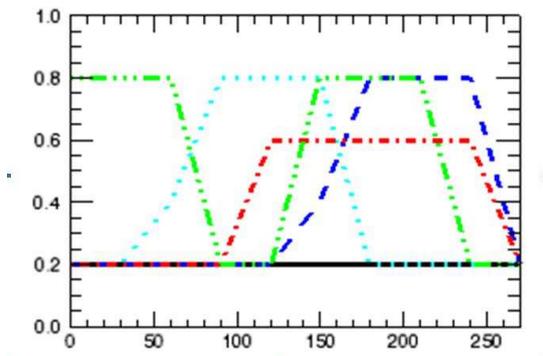}
\end{center}
\caption{\label{figure:Vindex} \it{The annual vegetation index. The
x-axis is the day of a year and different colors indicate different
land classes.}}
\end{figure}

The remote sensing image used in the experiment is about a cropland
with $5$ different land-use classes. The image size is $596$ pixel
by $529$ pixel. The features of interest are taken from the annual
vegetation index time series (see Figure~\ref{figure:Vindex}) at an
interval of $30$ days among which $10$ are used with each
corresponding to one scene of image at a different time of the year.
The vegetation index is an optical measure of vegetation canopy
'greenness' and is closely related to the photosynthetic potential
of plants. For each pixel, random noises, generated from Gaussian
$\mathcal{N}(0,0.1^2)$, are applied.

To simulate the acquisition of remote sensing images, the following
procedure is performed on each of the $10$ scenes of image.
\begin{enumerate}
\item
Rotate all images clockwisely by $10$ degrees.
\item
Re-sample each scene of image using a randomly generated offset from
$\mathcal{N}(0,0.1^2)$.
\item
Remove the blank edges in all images that are caused by rotation and
re-sampling.
\end{enumerate}
In Step 2 of the above, offsets are generated from the standard
Gaussian and a bilinear interpolation \cite{GomesDCV1998} is applied
during re-sampling. As a result, $247$ pixel by $233$ pixel
multi-temporal vegetation index images for the cropland of interest
are generated.

To assess the impact of image mis-registration to the task of
classification, two mis-registered images (corresponding to Case I
and II in Table~\ref{table:RSimage}, respectively) are generated
under different levels of mis-registration (roughly corresponding to
$3\%$ and $4\%$ data contamination, respectively). The SVM
classifier is trained on a sample from the original image and the
mis-registered image, respectively, and then test on a sample taken
from the original image. We use the data in a similar fashion as the
5-fold cross-validation, i.e., select $4$ folds for training and
rest for testing. Table \ref{table:RSimage} reports the
classification accuracy. We can see that, in both cases,
the loss in classification accuracy is small and can be well bounded
by our theoretical predication.
\begin{table}[ht]
\caption{\it{Accuracy of SVM for the cropland remote sensing image
under different amount of image mis-registration. Each of the first
$5$ columns corresponds to one of the $5$ folds. } }
\begin{center}
\begin{tabular}{|c||c|c|c|c|c|c|c|}
    \hline
   Fold        &1         &2        &3         &4       &5      &Average \\
     \hline
$Original$     &98.13     &98.14      &97.90    &97.94     &97.98    &98.02\\
\hline
Case I         &98.08     &98.11      &97.92    &97.94     &97.98    &98.01\\
\hline
Case II        &98.09     &98.10      &97.92    &97.92     &97.96    &97.99\\
\hline
\end{tabular}
\end{center}
\label{table:RSimage} 
\end{table}

It is known that, for example by bootstrap, the effect of
mis-registration on image classification varies with the relative
size of the ground area corresponding to an image pixel (call this
the pixel size) and the actual homogeneity (larger numbers
correspond to more homogeneity) of an area. If the ratio of these
two numbers is small, then the damage of mis-registration is small,
otherwise it is large. Since we are using a crop field here and the
corresponding pixel size is much smaller than that for the crop
field, the effect of data contamination is small. If the pixel size
is close to the actual object size, then mis-registration of half a
pixel may cause more damages.

\subsection{Some empirical results on Adaboost}
So far SVM has been used as the underlying classifier in our
experiment, other universally consistent classifiers such as
Adaboost are applicable as well. Instead of repeating the experiment
for AdaBoost, we collect results found in the literature
\cite{FreundSchapire1996,Dietterich1998,RF} and summarize in
Table~\ref{table:adaboost}. Note here we simply adopt the existing
results and this corresponds to taking $\epsilon=0.05$ only.
\begin{table}[htp]
\caption{\it{Error rates of Adaboost on some UC Irvine datasets
where 90\% of the data are used as the training set. Results are
shown for the original data and when 5\% of the class labels in the
training set are randomly flipped (uniformly into an alternate
class). Results are adopted from \cite{RF,
FreundSchapire1996,Dietterich1998} and then converted. }}
\begin{center}
\begin{tabular}{|c||c|c|c|}
    \hline
              &Original data     &5\% labels flipped    &Difference\\
    \hline
$Glass$           &22.00\% &22.35\%            &0.35\%\\
\hline
$Breast~cancer$   &~3.20\% &~4.58\%            &1.38\%\\
\hline
$Diabetes$      &26.60\% &28.41\%            &1.81\%\\
\hline
$Sonar$         &15.60\% &17.96\%            &2.36\%\\
\hline
$Ionsphere$      &~6.40\% &~8.17\%            &1.77\%\\
\hline
$Soybean$         &~7.57\% &~9.61\%            &2.04\%\\
\hline
$Ecoli$         &14.80\% &15.91\%            &1.11\%\\
\hline
$Votes$         &~4.80\% &~7.14\%            &2.34\%\\
\hline
$Liver$         &30.70\% &33.86\%            &3.16\%\\
\hline
\end{tabular}
\end{center}
\label{table:adaboost}
\end{table}

\subsection{Estimating the amount of data contamination}
\label{section:amtDC} Using data contamination bound
\eqref{eq:dataContBound2}, we can estimate the loss in accuracy for
classifiers trained with contaminated data. The remaining question
is to give a (rough) estimate of the amount of data contamination.
This is a question we would like to leave to future work.

In the special case of image mis-registration, we propose two simple
heuristics for estimating the amount of data contamination. Both are
based on the heuristic that the image pixels affected by
mis-registration are roughly those near the boundary between
different land classes. Thus the proportion of boundary pixels
serves as a good indication on the amount of data contamination.
Here the underlying assumption is that the proportion of boundary
pixels are roughly the same in the true and the mis-registered
images.

One approach is based on sampling. A number, say $100$ to $200$, of
pixels are randomly sampled from the image, we then count the
proportion of pixels that fall on the boundary by visual inspection.
Another estimate is based on the classification results by a
classifier trained on the contaminated data. For each pixel, we
determine if it is on the boundary by the following heuristic. For
each pixel in the image, take a $3 \times 3$ patch centering on it.
If there are at least two pixels within the patch having a different
class labels from the rest, then declare the pixel at the center of
the patch to be on the boundary. 
\section{Conclusion and discussion}
\label{section:conclusion} We formulate the problem of image
mis-registration as data contamination and equip it with a
statistical model. This model captures a very general class of
errors, for instance, measurement errors and gross errors that can
be formulated as label-flipping, feature-swapping, or feature
replacement by any proper distributions. Under a statistical
learning theoretical framework, we derive an asymptotic bound for
the loss in classification accuracy due to data contamination. One
nice feature about this bound is that, it is essentially
distribution-free thus it applies to all different types of data.
Extensive simulations on both synthetic and real datasets under
various types of data contaminations show that the data
contamination bound we derive is fairly tight. Compared to similar
bounds in the domain adaptation literature, our bound is sharper
and, unlike such bounds, our bound applies to classifiers with an
infinite VC dimension.

As we have already discussed, our data contamination model can
capture various types of errors such as image mis-registration,
label noise and accidental human errors. Beyond that, we can also
use data contamination as a useful device. We give here an example
in the setting of co-training (\cite{Yarowsky1995,
BlumMitchell1998,CollinsSinger1999}). Empirically, it has been shown
that co-training can significantly boost the classification accuracy
when the training sample size is extremely small, e.g., $12$ in
\cite{BlumMitchell1998} for web page classification and $6$ in
\cite{Nigam2000} for newsgroup classification. Theoretical work have
been carried out to understand the success of co-training (see, for
instance, \cite{BlumMitchell1998, DasguptaLM2001}). We provide here
a different perspective.

\begin{figure}[ht]
\centering
\begin{center}
\hspace{-0.15in}
\includegraphics*[scale=0.45,clip]{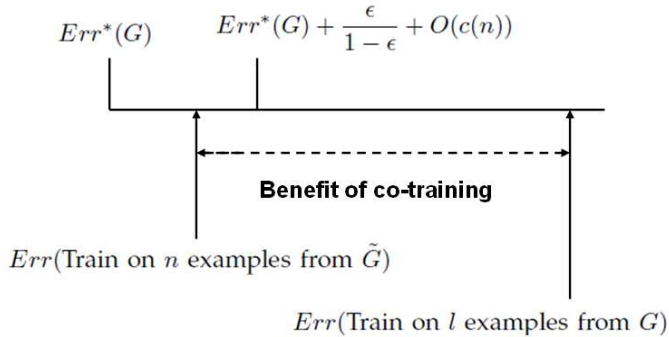}
\end{center}
\abovecaptionskip -15pt \caption{\it{The benefit of co-training.
$Err^*$ denotes the Bayes error rate.}} \label{figure:coTrainGap}
\end{figure}

In co-training, starting from a small amount of labeled examples,
the algorithm progressively enlarges the labeled set by transferring
those examples which are originally unlabeled but are classified
with high confidence by the classifier built from the labeled data
available so far. This amounts to enlarging the labeled set with a
small amount of label noise; the label noise here is small because
those examples which are being transferred are classified with high
confidence. Assume at certain point we have $n$ examples in the
labeled set and assume $n$ is large, then, by our analysis (c.f.
\eqref{eq:dataContBound2}), the additional classification error
w.r.t. that resulting from a clean labeled set (of size $n$) is no
more than $\epsilon/(1-\epsilon)+O(c(n))$ for $c(n) \rightarrow 0$
as $n$ grows. Thus,
\begin{eqnarray*}
&& Err(\mbox{Bayes classifier on $G$}) \\
&\leq& Err(\mbox{Classifier learned on $n$ observations from $\tilde{G}$}) \\
&\leq&  Err(\mbox{Bayes classifier on
$G$})+\frac{\epsilon}{1-\epsilon}+O\left(c(n)\right)
\end{eqnarray*}
where $Err$ denotes the error rate. Here, we use $G$ and $\tilde{G}$
to denote the data with clean label and that containing labels
assigned by the co-training algorithm, respectively. It is clear
that the error rate achieved by co-training equals that by a
classifier learned on $n$ observations from $\tilde{G}$. However, it
is often the case that the error rate by a classifier learned on $l$
labeled examples from $G$ is typically much larger, i.e.,
\begin{eqnarray}
&& Err(\mbox{Classifier learned on $l$ examples from $G$}) \nonumber
\\
&\gg& Err(\mbox{Bayes classifier on
$G$})+\frac{\epsilon}{1-\epsilon}+O\left(c(n)\right)
\label{eq:gapCotrain}
\end{eqnarray}
if $l$ is small, $\epsilon$ is small and $n$ is large. The gap
between the two $Err$ terms in \eqref{eq:gapCotrain} is the
potential ``benefit'' of co-training as illustrated in
Figure~\ref{figure:coTrainGap}. This explains why co-training may be
feasible with a small amount of initial labeled examples. Since the
gap in \eqref{eq:gapCotrain} shrinks as $l$ increases, this, on the
other hand, explains why co-training may not help much when the
initial labeled set is large.

A limitation of our data contamination model~\eqref{eq:DCModel} is
that, in modeling the phenomenon of image mis-registration with a
data contamination model, i.i.d. contaminations are assumed.
However, in practice the mis-registered image pixels may be
correlated in some way. It is thus desirable to take this into
account in the model, which we shall leave to future work.
Note that we derive the data contamination bound under a general
class of data distributions, it is desired to take advantage of
knowledge on the underlying distribution to get a sharper bound.
Note also that the focus of the present paper is the analysis and
simulation on the impact of data contamination to classification
accuracy, no new algorithm is proposed. We shall leave that to
future work, interested readers can see, for example,
\cite{ShivaswamyBhattacharyyaSmola2006} and references therein.

\section*{Acknowledgments}

The authors would like to thank Tin Kam Ho at Bell Labs for kindly
providing the four-class and nested-square datasets.


\bibliographystyle{plain}
\bibliography{./cdc}

\begin{thebibliography}{10}

\bibitem{UCI}
A.~Asuncion and D.~J. Newman.
\newblock {UCI Machine Learning Repository, Department of Information and
  Computer Science}.
\newblock {http://www.ics.uci.edu/~mlearn/MLRepository.html}, 2007.

\bibitem{BartlettMendelson2001}
P.~L. Bartlett and S.~Mendelson.
\newblock Rademacher and gaussian complexities: Risk bounds and structural
  results.
\newblock In {\em COLT}, pages 224--240, 2001.

\bibitem{BartlettAmbuj2007}
P.~L. Bartlett and A.~Tewari.
\newblock Sparseness vs estimating conditional probabilities: Some asymptotic
  results.
\newblock {\em Journal of Machine Learning Research}, 8:775--790, 2007.

\bibitem{BartlettTraskin2007}
P.~L. Bartlett and M.~Traskin.
\newblock Adaboost is consistent.
\newblock {\em Journal of Machine Learning Research}, 8:2347--2368, 2007.

\bibitem{Ben-DavidBCKPV2010}
S.~Ben-David, J.~Blitzer, K.~Crammer, A.~Kulesza, F.~Pereira, and J.~Wortman
  Vaughan.
\newblock A theory of learning from different domains.
\newblock {\em Machine Learning}, 79:151--175, 2010.

\bibitem{BlumMitchell1998}
A.~Blum and T.~Mitchell.
\newblock Combining labeled and unlabeled data with co-training.
\newblock In {\em Proceedings of the eleventh annual conference on
  Computational learning theory}, pages 92--100, 1998.

\bibitem{Bagging}
L.~Breiman.
\newblock Bagging predicators.
\newblock {\em Machine Learning}, 24(2):123--140, 1996.

\bibitem{RF}
L.~Breiman.
\newblock Random {F}orests.
\newblock {\em Machine Learning}, 45(1):5--32, 2001.

\bibitem{carrollDelaige2009}
R.~J. Carroll, A.~Delaigle, and P.~Hall.
\newblock Nonparametric prediction in measurement error models.
\newblock {\em Journal of the American Statistical Association}, 2009 (To
  appear).

\bibitem{LIBSVM}
C.-C. Chang and C.-J. Lin.
\newblock {\em {LIBSVM}: a library for support vector machines}, 2001.
\newblock Software available at {http://www.csie.ntu.edu.tw/~cjlin/libsvm}.

\bibitem{CollinsSinger1999}
Michael Collins and Yoram Singer.
\newblock Unsupervised models for named entity classification.
\newblock In {\em In Proceedings of the Joint SIGDAT Conference on Empirical
  Methods in Natural Language Processing and Very Large Corpora}, pages
  100--110, 1999.

\bibitem{DasguptaLM2001}
S.~Dasgupta, M.L. Littman, and D.~McAllester.
\newblock {PAC} generalization bounds for co-training.
\newblock In {\em Proceedings of Neural Information Processing Systems (NIPS)},
  pages 375--382, 2001.

\bibitem{DefriesTownshend1999}
R.~S. Defries and J.~R.~G. Townshend.
\newblock Global land cover characterization from satellite data: from research
  to operational implementation?
\newblock {\em Global Ecology and Biogeography}, 8(5):367--379, 1999.

\bibitem{DelaigeFan2009}
A.~Delaigle, J.~Fan, and R.~J. Carroll.
\newblock A design-adaptive local polynomial estimator for the
  errors-in-variables problem.
\newblock {\em Journal of the American Statistical Association}, 104:348--359,
  2009.

\bibitem{DevroyeGyorfiLugosi1996}
L.~Devroye, L.~Gy\"{o}rfi, and G.~Lugosi.
\newblock {\em A Probabilistic Theory of Pattern Recognition (Stochastic
  Modelling and Applied Probability)}.
\newblock Springer, 1996.

\bibitem{Dietterich1998}
T.~G. Dietterich.
\newblock An experimental comparison of three methods for constructing
  ensembles of decision trees: Bagging, boosting and randomization.
\newblock {\em Machine Learning}, 40(2):139--157, 1998.

\bibitem{FreundSchapire1996}
Y.~Freund and R.~E. Schapire.
\newblock Experiments with a new boosting algorithm.
\newblock In {\em Proceedings of the 13rd International Conference on Machine
  Learning (ICML)}, 1996.

\bibitem{Fuller1987}
W.~A. Fuller.
\newblock {\em Measurement Error Models}.
\newblock John Wiley, 1987.

\bibitem{GomesDCV1998}
J.~Gomes, L.~Darsa, B.~Costa, and L.~Velho.
\newblock {\em Warping and Morphing of Graphical Objects}.
\newblock Morgan Kaufmann, 1998.

\bibitem{GongLeDrewMiller1992}
P.~Gong, E.~F. LeDrew, and J.~R. Miller.
\newblock Registration noise reduction in difference images for change
  detection.
\newblock {\em International Journal of Remote Sensing}, 13(4):773--779, 1992.

\bibitem{GongXu2003}
P.~Gong and B.~Xu.
\newblock Remote sensing of forests over time: change types, methods, and
  opportunities.
\newblock In M.~Woulder and S.~E. Franklin, editors, {\em Remote Sensing of
  Forest Environments: Concepts and Case Studies}, pages 301--333. Kluwer
  Press, Amsterdam, Netherlands, 2003.

\bibitem{Hampel1974}
F.~R. Hampel.
\newblock The influence curve and its role in robust estimation.
\newblock {\em Journal of the American Statistical Association},
  69(346):383--393, 1974.

\bibitem{HoKleinberg1996}
T.~K. Ho and E.~M. Kleinberg.
\newblock Building projectable classifiers of arbitrary complexity.
\newblock In {\em International Conference on Pattern Recognition}, 1996.

\bibitem{Huber1972}
P.~J. Huber.
\newblock Robust statistics: A review ({T}he 1972 {W}ald {L}ecture).
\newblock {\em The Annals of Mathematical Statistics}, 43(4):1041--1067, 1972.

\bibitem{Jensen2004}
J.~R. Jensen.
\newblock {\em Introductory Digital Image Processing}.
\newblock Prentice Hall, 2004.

\bibitem{JusticeVermoteTownshend1998}
C.~O. Justice, E.~Vermote, J.~R.~G. Townshend, R.~Defries, D.~P. Roy, D.~K.
  Hall, V.~V. Salomonson, J.~L. Privette, G.~Riggs, A.~Strahler, W.~Lucht,
  R.~B. Myneni, Y.~Knyazikhin, S.~W. Running, R.~R. Nemani, Z.~Wan, A.~R.
  Huete, W.~van Leeuwen, R.~E. Wolfe, L.~Giglio, J.~Muller, P.~Lewis, and M.~J.
  Barnsley.
\newblock The moderate resolution imaging spectroradiometer ({MODIS}): Land
  remote sensing for global change research.
\newblock {\em IEEE Transactions on Geoscience and Remote Sensing},
  36(4):1228--1249, 1998.

\bibitem{LiuKellyGong2006}
D.~Liu, M.~Kelly, and P.~Gong.
\newblock A spatio-temporal approach to monitoring forest disease spread using
  multi-temporal high spatial resolution imagery.
\newblock {\em Remote Sensing of Environment}, 101(2):167--180, 2006.

\bibitem{MansourMohriRostamizadeh2009}
Y.~Mansour, M.~Mohri, and A.~Rostamizadeh.
\newblock Domain adaptation: Learning bounds and algorithms.
\newblock In {\em COLT}, 2009.

\bibitem{Nigam2000}
K.~Nigam.
\newblock Understanding the behavior of co-training.
\newblock In {\em In Proceedings of KDD-2000 Workshop on Text Mining}, 2000.

\bibitem{Quinlan1986}
J.~Quinlan.
\newblock The effect of noise on concept learning.
\newblock In R.~S. Michalski, J.~G. Carbonell, and T.~M. Mitchell, editors,
  {\em Machine Learning, an Artificial Intelligence Approach, Volume II}, pages
  149--166. Morgan Kaufmann, 1986.

\bibitem{ShivaswamyBhattacharyyaSmola2006}
P.~K. Shivaswamy, C.~Bhattacharyya, and A.~J. Smola.
\newblock Second order cone programming approaches for handling missing and
  uncertain data.
\newblock {\em Journal of Machine Learning Research}, 7:1283--1314, 2006.

\bibitem{Sloan1988}
R.~H. Sloan.
\newblock Types of noise in data for concept learning.
\newblock In {\em The First Workshop on Computational Learning Theory}, pages
  91--96. Morgan Kaufmann, 1988.

\bibitem{Steinwart2002}
I.~Steinwart.
\newblock Support vector machines are universally consistent.
\newblock {\em Journal of Complexity}, 18:768--791, 2002.

\bibitem{Steinwart2005}
I.~Steinwart.
\newblock Consistency of support vector machines and other regularized kernel
  machines.
\newblock {\em IEEE Transactions on Information Theory}, 51:128--142, 2005.

\bibitem{SwainVanderbilt1982}
P.~H. Swain, V.~C. Vanderbilt, and C.~D. Jobusch.
\newblock A quantitative applications-oriented evaluation of thematic mapper
  design specifications.
\newblock {\em IEEE Transactions on Geoscience and Remote Sensing},
  20(3):370--377, 1982.

\bibitem{TownshendJusticeGurney1992}
J.~R.~G. Townshend, C.~O. Justice, C.~Gurney, and J.~McManus.
\newblock The impact of mis-registration on change detection.
\newblock {\em IEEE Transactions on Geoscience and Remote Sensing},
  30(5):1054--1060, 1992.

\bibitem{Tukey1960}
J.~W. Tukey.
\newblock A survey of sampling from contaminated distributions.
\newblock In I.~Olkin, editor, {\em Contributions to Probability and
  Statistics}, pages 448--485. Standford University Press, 1960.

\bibitem{Vapnik1998}
V.~N. Vapnik.
\newblock {\em Statistical Learning Theory}.
\newblock John Wiley, 1998.

\bibitem{XuDickson2009}
Y.~Xu, B.~G. Dickson, H.~M. Hampton, T.~D. Sisk, J.~A. Palumbo, and J.~W.
  Prather.
\newblock Effects of mismatches of scale and location between predictor and
  response variables on forest structure mapping.
\newblock {\em Photogrammetric Engineering \& Remote Sensing}, 75(3):313--322,
  2009.

\bibitem{Yarowsky1995}
David Yarowsky.
\newblock Unsupervised word sense disambiguation rivaling supervised methods.
\newblock In {\em in Proceedings of the 33rd Annual Meeting of the Association
  for Computational Linguistics}, pages 189--196, 1995.

\bibitem{ZhuWu2004}
X.~Zhu and X.~Wu.
\newblock Class noise vs. attribute noise: a quantitative study of their
  impacts.
\newblock {\em Artificial Intelligence Review}, 22(3):177--210, 2004.

\end{thebibliography}

\end{document}